\documentclass[default]{aastex63}
\newcommand{\hl}[1]{#1}
\usepackage{chemformula}

\submitjournal{PSJ}

\shorttitle{Silica in martian fans and deltas}
\shortauthors{Pan et al.}

\graphicspath{{./}{./}}

\begin{document}
\title{Voluminous silica precipitated from \hl{martian waters during late-stage aqueous alteration}}

\correspondingauthor{Lu Pan}
\email{lu.pan@sund.ku.dk}
\author[0000-0002-8151-2125]{Lu Pan}
\affiliation{Univ. Lyon, Univ. Lyon 1, ENS Lyon, CNRS, LGL-TPE, F-69622, Villeurbanne, France}
\affiliation{University of Copenhagen, GLOBE institute, Centre for Star and Planet Formation, Copenhagen, Denmark}

\author[0000-0002-2698-6926]{John Carter}
\affiliation{Institut d’Astrophysique Spatiale, CNRS, Université Paris-Sud, Orsay, France}

\author[0000-0002-8313-8595]{Cathy Quantin-Nataf}
\affiliation{Univ. Lyon, Univ. Lyon 1, ENS Lyon, CNRS, LGL-TPE, F-69622, Villeurbanne, France}

\author[0000-0002-1439-8131]{Maxime Pineau}
\affiliation{Université de Nantes, Laboratore de Planétologie et Géodynamique (LPG),  UMR-CNRS 6112, Nantes, France}

\author[0000-0003-0431-6312]{Boris Chauviré}
\affiliation{Université Grenoble-Alpes, CNRS, Grenoble, France}

\author[0000-0002-0022-0631]{Nicolas Mangold}
\affiliation{Université de Nantes, Laboratore de Planétologie et Géodynamique (LPG),  UMR-CNRS 6112, Nantes, France}

\author[0000-0003-1361-5170]{Laetitia Le Deit}
\affiliation{Université de Nantes, Laboratore de Planétologie et Géodynamique (LPG),  UMR-CNRS 6112, Nantes, France}

\author[0000-0003-2954-2770]{Benjamin Rondeau}
\affiliation{Université de Nantes, Laboratore de Planétologie et Géodynamique (LPG),  UMR-CNRS 6112, Nantes, France}

\author[0000-0002-1111-587X]{Vincent Chevrier}
\affiliation{University of Arkansas, Arkansas Center for Space and Planetary Sciences, USA}

\begin{abstract}
\hl{Mars’ transition from an early “warm and wet” to the “cold and dry” environment left fingerprints on the geological record of fluvial activity on Mars.
The morphological and mineralogical observations of aqueous activity provided varying constraints on the condition and duration of liquid water on martian surface.
In this study, we surveyed the mineralogy of martian alluvial fans and deltas and investigated the hydrated silica-bearing deposits associated with these landforms.
Using CRISM data, we identified 35 locations across Mars with hydrated silica in proximity to  fan/deltas, where the spectral characteristics are consistent with immature or dehydrated opal-A.
In a few stepped fan/deltas, we find hydrated silica occurs within the bulk fan deposits and form sedimentary layers correlated with elevation, corroborating  the  formation  of  hydrated  silica  through  precipitation.
Meanwhile in the older fan/deltas silica mostly  occur at distal locations and the relation to primary sedimentary deposits is more complex.
We propose that the hydrated silica-bearing deposits in stepped fan/deltas likely formed authigenically from martian surface waters, mainly during the Late Hesperian and Early Amazonian \citep{hauber_asynchronous_2013}.
These silica-bearing deposits could be a tracer for the temperature of water involved in the formation of these deposits, given more precise and detailed observations of the sedimentary context, accessory minerals, the concentration of hydrated silica and sediment-to-water ratio.
Therefore, we consider that silica-bearing deposits should be among the most critical samples to investigate for future Mars missions, which are accessible in the landing sites of Mars 2020 and ExoMars missions.}
\end{abstract}


\section{Introduction}\label{sec:intro}
Liquid water \hl{flowed} on the surface of Mars before 3.5 billion years ago, building a broad diversity of aqueous and potentially habitable environments.
The valley networks, deltas, alluvial fans, paleolakes observed from orbit \citep[e.g.,][]{mangold_evidence_2004,kite_persistent_2017} and \textit{in situ} \hl{by landed mission} \citep[e.g.,][]{grotzinger_deposition_2015} suggest the presence of long-lasting continued runoff and large volumes of liquid water on the surface of Mars that \hl{were most prevalent during the Noachian and Hesperian periods, but in some cases extended to Late Hesperian and Amazonian (e.g., dendritic networks in Valles Marineris \citep{mangold_evidence_2004}).
Conversely, the mineralogy record of past aqueous activity preserved immature minerals (e.g., smectites, opaline silica), whose survival until today point to limited water-rock interaction and diagenesis after their formation \citep{tosca_juvenile_2009}.}
To reconcile these different observations and better understand the climate of Mars through time, we need to reexamine the duration, timing, and physical states \hl{of past aqueous environments on Mars}.
Here, our investigation focuses on the mineralogy of the sedimentary deposits of surface water activity on Mars, specifically the hydrated silica associated with the fans and deltas, to understand the evolutionary pathway of the martian climate through time.

Hydrated silica (\ch{SiO_2.nH_2O}) is a common weathering product of basaltic rocks, the main constituent of Mars' surface \citep{ shoji_opaline_1971,ping_properties_1988,opfergelt_silicon_2011,mclennan_sedimentary_2003}.
Hydrated silica has been identified on Mars from orbit \citep[e.g.][]{milliken_opaline_2008,bishop_phyllosilicate_2008,bandfield_high-silica_2008,skok_silica_2010,goudge_constraints_2012,carter_hydrous_2013,smith_hydrated_2013,hauber_asynchronous_2013,sun_distinct_2018,pineau_toward_2020} and \textit{in situ} \hl{with landed missions} \citep[e.g.,][]{squyres_detection_2008,rice_silica-rich_2010,rapin_situ_2018}.
More recently, \hl{silica-bearing deposits} have been found in the Mars2020 landing site Jezero Crater \citep{tarnas_orbital_2019}, as well as Oxia Planum, the selected landing site for European Space Agency’s ExoMars rover \citep{quantin-nataf_exomars_2019,carter_oxia_2016}.
\hl{The precipitation of silica occurs when a solution becomes oversaturated with respect to the mineral phase (quartz, amorphous silica) as it reaches certain physical and chemical conditions} \citep[e.g.,][]{alexander_solubility_1954,siever_silica_1962,sjoberg_silica_1996}.
Thus, \hl{silica-bearing deposits can be an essential indicator of the temperature and to a lesser extent, pH of the aqueous environment in which they are formed.}
In addition, various forms of hydrated silica (and their diagenetic transformation products) have been characterized as high astrobiological potential, due to their ability to entrap and preserve \hl{microfossils and other chemical and morphological biosignatures} \citep{Knoll_Whittington_Morris_1985,trewin1996rhynie,lazzeri_nitrogen_2017,mcmahon_field_2018,teece_biomolecules_2020,abu-mahfouz_silica_2020}.
Locating and investigating the \hl{silica-bearing deposits} in fluvial settings will thus bring new understanding to the climate conditions in which fluvial morphology and silica co-occurred and highlight important scientific targets for future \textit{in situ} observations.

On Mars, \hl{terminal fluvial} deposits have been identified to be widespread on the interior of impact craters that are connected to incised channels  \citep[e.g.,][]{cabrol_distribution_1999,irwin_sedimentary_2004, moore_large_2005,di_achille_ancient_2010}.
Many of these deposits are interpreted as alluvial fans, formed by sediment deposition into a conical shape due to lateral expansion of flow   \citep[e.g.,][]{moore_large_2005,kraal_catalogue_2008}, while others show \hl{inverted channel avulsion on the surface that are interpreted as branched deltas}  \citep[e.g.,][]{pondrelli_complex_2005,fassett_fluvial_2005}.
On Mars, additionally, a type of "stair-step" or "terraced" fan-shaped sediment, which lacks typical terrestrial analog, has been proposed \hl{to form due to one of these scenarios:} a deltaic deposition with increasing water base level  \citep[e.g.,][]{de_villiers_experimental_2013}; erosional wave actions \citep[e.g.,][]{ori_terraces_2000}; or \hl{multi-episode} debris-flow dominated alluvial processes \citep{di_achille_steep_2006}.
Here we study \hl{silica-bearing deposits} associated with alluvial fans and deltas on Mars using a combination of available orbital image datasets.
Based on the geological context of the hydrated \hl{silica-bearing deposits}, we discuss their possible formation scenarios and their implications for the martian environment.
Our analysis aims at providing an overview to guide the investigation of \hl{silica-bearing deposits} in alluvial fans and deltas during future \textit{in situ} explorations and Mars Sample Return missions, which could bring crucial constraints on the conditions (volume, temperature, pH and mass concentration) of past martian waters.

\section{Method} \label{sec:method}
\subsection{A global survey of silica-bearing fans and deltas}
    The Compact Reconnaissance Imaging Spectrometer for Mars (CRISM) is a hyperspectral imaging spectrometer that acquires data of the martian surface in the near infrared wavelength range (1-4 $\micron$), which enables the identification of hydrated minerals, including hydrated silica \citep{murchie_compact_2007,murchie_synthesis_2009}.
    We built our global datasets based on a selection of CRISM images from two sources: (i) locations from a previous global survey of CRISM images \citep{carter_hydrous_2013} filtered by morphology indicators as alluvial fans and deltas; (ii) a database of alluvial fans and deltas identified in previous morphology analysis \citep[e.g.,][]{irwin_intense_2005,cabrol_distribution_1999,cabrol_evolution_2001, ori_terraces_2000,moore_martian_2003,kraal_catalogue_2008,di_achille_ancient_2010} (Table S1) where we confirmed CRISM coverage in proximity to these deposits.
    The complete list of the fans and deltas investigated in this study with required CRISM coverage is given in Table S2, with
    relevant references where available.

    We analyzed the select CRISM images following the previous data processing methodology, in which we calculated the ratioed images based on column-averaged denominator of selected pixels below the 1.5-sigma threshold within the spectral parameter range \citep{carter_automated_2013,pan_stratigraphy_2017}.
    Hydrated silica was identified based on absorption bands at 1.4, 1.9 $\micron$ and a broad, characteristic 2.2 $\micron$ band due to vibrational modes of hydroxyl groups in molecular water (\ch{H_2O}) and silanol groups (\ch{Si-OH}) \hl{\citep{anderson_near_1964,langer_near_1974}}.
    The averaged spectra were obtained from a region of interest within contiguous pixels exhibiting a similar NIR reflectance value and texture in CRISM images (Figure \ref{fig:A1}).
    The spectra with possible identification of hydrated silica were classified based on \hl{the confidence level in spectral feature identification}, quantified by the 1.9 and 2.2 $\micron$ band depths (Table S3, Figure \ref{fig:A2}).
    The classification only applied to the type spectra chosen for each location, while spectral features may vary within a given fan/delta location.
\subsection{High-resolution images and Digital Elevation Model (DEM)}
    We utilized imagery datasets of various spatial resolutions to investigate the silica-bearing deposits.
    We used the CTX mosaic provided by the Murray Planetary Visualization Lab as a base map \citep{dickson_global_2018}.
    The High Resolution Imaging Experiment (HiRISE) ($\sim$25 cm/pix) \citep{mcewen_mars_2007} and Context Camera (CTX) ($\sim$6m/pix) \citep{malin_context_2007} images of locations of interest were downloaded and calibrated using the \hl{Integrated Software for Imagers and Spectrometers (ISIS)} routine enabled by the MarsSI platform \citep{quantin-nataf_marssi:_2018}.
    All the visible images and CRISM spectral data with various resolutions were co-registered to the CTX mosaic using the georeferencing application of the Quantum GIS (QGIS) software and manually selecting the control points.
    The outlines of the fans were digitized tracing the base of the fans also within a QGIS project.
    For the key sites with stereo image coverage, we processed the stereo images to generate DEMs with USGS Ames pipeline \citep{beyer_ames_2018}.
    Topographic profiles, elevation, slope and the maximum fan volumes were measured for locations with clear morphologic context.
    We estimated the volume of the fan deposit with a flat-bottom assumption, which likely overestimates the total volume of the deposit.
\subsection{Spectral analysis of hydrated silica}
    The free and hydrogen-bonded water species (\ch{H_2O} and \ch{Si-OH}) within hydrated silica vary in different types of silica due to their varying crystallinity and formation pathways \citep[e.g.,][]{anderson_near_1964,langer_near_1974,bobon_state_2011,rice_reflectance_2013,chauvire_near_2017}.
    To understand the type of silica deposited with fans and deltas, we calculated spectral parameters as demonstrated in previous studies \citep{rice_reflectance_2013,sun_distinct_2018,pineau_toward_2020}.
    It has been shown that the most distinguishing spectral parameters are the location of the minimum of the 1.4 $\micron$ band and the shape of the 1.9 and 2.2 $\micron$ bands.
    The minimum of the 1.4 $\micron$ band is calculated upon the linear continuum removed spectra anchored at 1.35 and 1.44 $\micron$.
    We calculated the relative band depths using several different approaches.
    Following \citet{rice_reflectance_2013}, the 1.91 and 1.96 $\micron$ band depths are calculated using the classic band depth formula (Eq. \ref{eq:banddepth}).
    \begin{equation}\label{eq:banddepth}
        BD= 1-\frac{R_b}{R_c}
    \end{equation}
    Where $R_b$ is the reflectance of the band position and $R_c$ is the reflectance calculated for the continuum made up of both shoulders of the absorption band.
    The band ratio of 2.21 and 2.26 $\micron$ were calculated by fitting a double gaussian centered at $\sim$2.21 and $\sim$2.26 $\micron$ on the continuum removed spectra, equivalent to the method of Sun and Milliken [2018].
    Finally, we applied the new spectral parameters (CRC1.4 and CRC1.9) based on terrestrial opals, which allows discerning hydrothermal and low-temperature weathering hydrated silica and has been shown to be \hl{potentially} applicable to Mars \citep{chauvire_near_2017,pineau_toward_2020}.
    The spectra were first smoothed using the Savitzky-Golay algorithm with third-degree polynomial functions and a moving average windows of about 7 to 9 spectels \citep{savitzky_smoothing_1964,steinier_smoothing_1972}.
    For the two absorption bands at 1.4 and 1.9 $\micron$, two anchor points were chosen at $\sim$1.30 and $\sim$1.60 $\micron$, and $\sim$1.85 and $\sim$2.10 $\micron$ respectively.
    Then, we calculated the two band-depth ratios: BDR, the real band-depth ratio of the inflexions points at 1.46 and 1.96 $\micron$ (Eq. \ref{eq:crc}), and $BDR_{*}$, the virtual band-depth ratio of the same wavelength using the linear continuum between the minimum of 1.4/1.9 $\micron$ band and the anchor point at $\sim$1.60 $\micron$, $\sim$2.10 $\micron$, respectively.
    Eventually, the CRC parameters were calculated as:
    \begin{equation}\label{eq:crc}
        CRC =  \frac{BDR}{BDR_{*}}
    \end{equation}
    The errors in the CRC calculations were calculated empirically by shifting the two anchor points over +/- 3 spectels.
    As a result, several CRC were calculated for each absorption band of each spectrum (over 35 combinations).
    The CRC retained value is the average of all the calculated values, and the errors represent the standard deviation ($1 \sigma$).
    The more convex the absorption band is, the higher the CRC value is.
    \hl{CRC1.9 is highly sensitive to the atmospheric residual at 1.96-2.1 $\micron$ due to \ch{CO_2} absorption and CRC1.4 can be affected when absorption is small in comparison with the noise.
    Therefore we removed spectra with particularly weak 1.4 and 1.9 $\micron$ bands from this analysis, and plotted with errorbars calculated using shifted anchor points to provide a first-order evaluation of the effect of noise. A full dataset of calculated CRC parameters can be found in Figure \ref{fig:A3}.}

\begin{figure}[htb!]
    \centering
    \includegraphics[width=\textwidth]{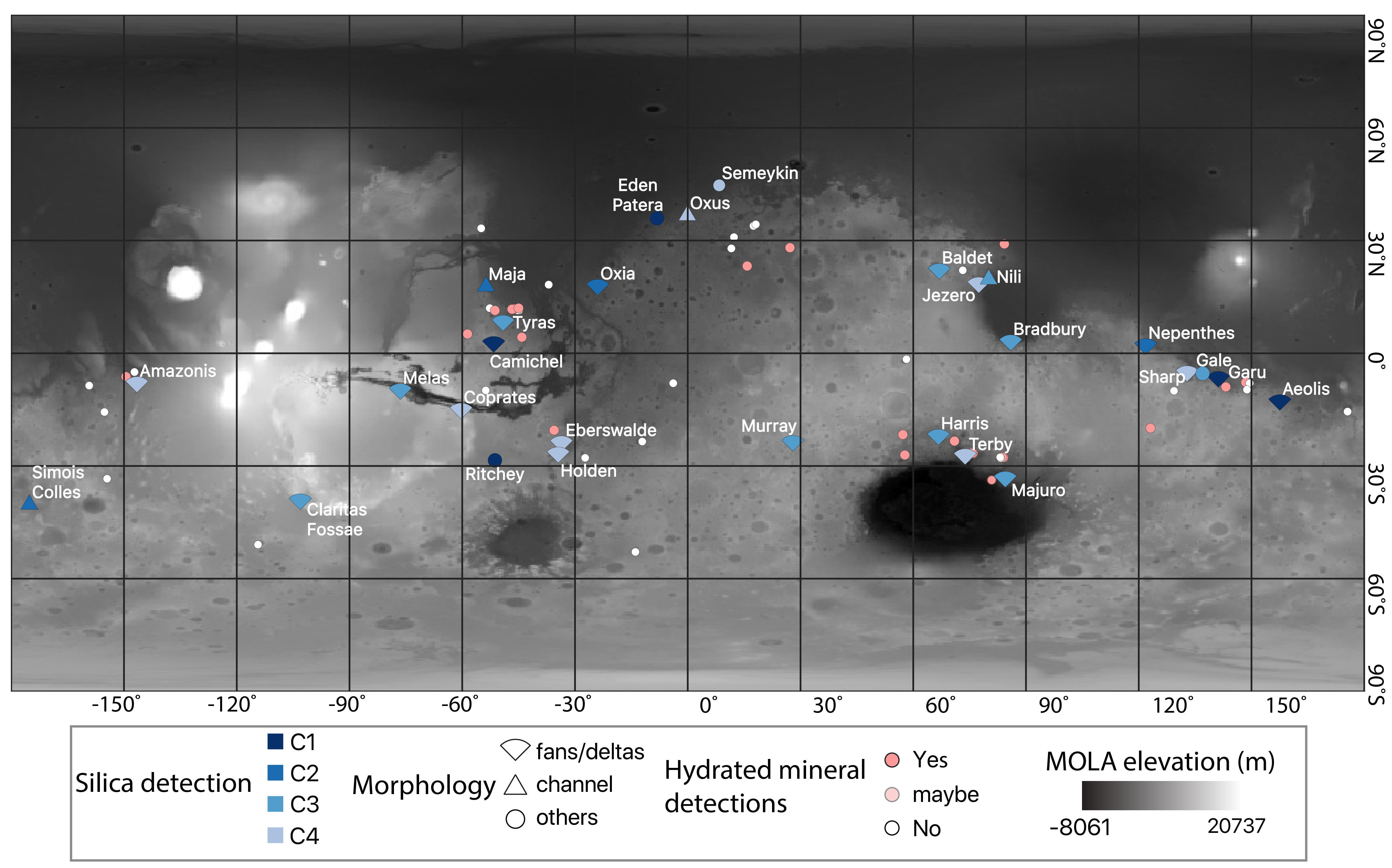}
    \caption{Global distribution of fluvial deposits associated with hydrated \hl{silica-bearing deposits}.
    This map highlights the locations with fluvial morphology found with hydrated minerals in a previous global survey \citep{carter_hydrous_2013}.
    Locations found with hydrated silica detections are highlighted in blue color and those without a clear silica detection in pink. The silica spectral signatures of different confidence are categorized into C1-C4 based on their band depths at 1.9 and 2.2 $\micron$ (See Section 2.1).
    For locations found with \hl{silica-bearing deposits}, fan symbols are used to represent \hl{silica-bearing deposits} associated with fans or deltas; triangles represent silica found associated with deposits within a fluvial channel; dots represent other types of geological settings (e.g., a volcanic caldera, periglacial setting).
    The background is MOLA in equirectangular projection.}
    \label{fig:global}
\end{figure}

\section{Results: Mineralogy, distribution and context for \hl{silica-bearing deposits}}\label{sec:results}
Within the 98 locations of previously identified alluvial fans or delta with CRISM coverage (Table S1), we identify 62 with confirmed or possible identification of hydrated minerals.
Among them, 35 locations (Figure \ref{fig:global}) with associated fluvial deposits are found with hydrated \hl{silica-bearing deposits}, including the two future landing sites at Jezero crater and Oxia Planum, for Mars2020 and ExoMars2022 missions, respectively.
Since the identification of hydrated minerals using CRISM images can be confounded by dust cover, partial coverage of the fan/delta and exposures beneath the spatial resolution of the images ($\sim$18-36 m/pixel), the number of identified locations gives a lower limit to the global distribution of hydrated silica occurrences in fans and deltas.
Regardless, hydrated silica, which is rather spatially limited compared to other hydrous minerals on Mars (e.g., Fe/Mg phyllosilicates) \citep[e.g.,][]{carter_hydrous_2013,ehlmann_mineralogy_2014}, is found to occur \hl{relatively} frequently in alluvial fans and deltas.
All these \hl{silica-bearing deposits} identified are observed within the spatial extent of sedimentary basins \hl{that have been affected by fluvial activity.}
Most of the fluvial deposits form distinct morphology of alluvial fans or deltas, while others have preserved evidence for a paleolake  (e.g. Melas Chasma \citep{metz_sublacustrine_2009,weitz_mixtures_2015}) and at proximity to fluvial landforms like channels (e.g. Maja Valles).
\hl{Silica-bearing deposits} found within the Maja Valles channel are also at the terminus of smaller, better-preserved fluvial channels, but no clear fan/delta morphology is identified (Figure \ref{fig:A4}).
\begin{figure}[htb!]
    \centering
    \includegraphics[width=0.95\textwidth]{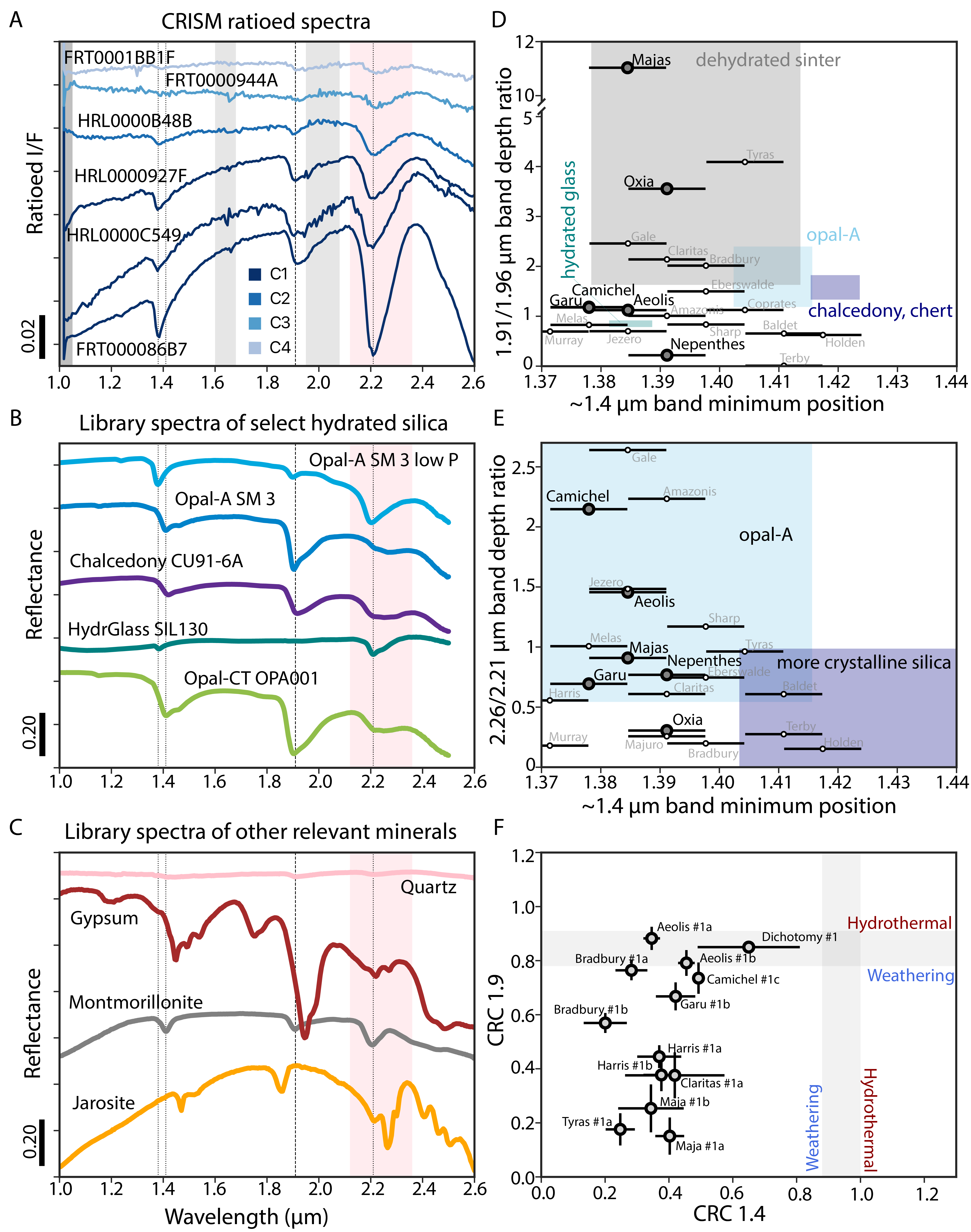}
    \hl{
    \caption{(Caption next page.)}}
\end{figure}
\addtocounter{figure}{-1}
\begin{figure}
  \caption{(Previous page.) Spectral detections of hydrated silica on Mars. A. Select CRISM spectra of hydrated silica detections with varying certainties in alluvial fans. \hl{The grey shaded regions represent wavelengths of CRISM artifacts and atmospheric residuals. The pink shaded region corresponds to the typical 2.2 $\micron$ band of hydrated silica.} B. Type laboratory spectra of different varieties of hydrated silica found on Earth. Opal-CT and hydrated glass spectra are from \citet{rice_reflectance_2013}; Opal-A spectra under normal and low-pressure conditions are from \citet{sun_distinct_2018}. The spectrum for chalcedony is from USGS spectral library \citep{clark_usgs_2007}. C. Relevant mineral spectra for comparison from spectral libraries, including quartz, Al-phyllosilicate, and sulfate  \citep{clark_usgs_2007}. D. Plot of 1.4 $\micron$ band minimum position and 1.91/1.96 $\micron$ band depth ratio for select locations (excluding locations with ambiguous context and where the 1.9 $\micron$ band is strongly affected by the atmosphere residual). Larger grey dots represent spectra categorizes as C1 or C2. The ranges of different types of silica are from \citet{rice_reflectance_2013}. E. Plot of spectral absorption of 1.4 $\micron$ band position and 2.21/2.26 $\micron$ band shape following \citet{sun_distinct_2018}. Larger grey dots represent spectra categorizes as C1 or C2. \hl{Blue shaded regions represent the parameter space of opal-A and opal-C/CT respectively.} F. The plot of CRC1.9 and CRC1.4 spectral parameters compared to the range of different types of terrestrial silica samples following \citet{pineau_toward_2020}. A larger set of spectra (1-4) for each location has been examined. CRC parameters that are affected by atmosphere residual or low signal-to-noise level are not included.}
  \label{fig:spec}
\end{figure}

We identify hydrated mineral detections using CRISM spectra with absorption bands around 1.4 $\micron$, 1.9 $\micron$ and a broad 2.2 $\micron$ due to vibrational modes of hydroxyl groups in molecular water (\ch{H_2O}) and silanol groupments \ch{Si-OH} (Figure \ref{fig:spec}A).
Although there are variabilities in the band position and band shapes of 1.4, 1.9, and 2.2 $\micron$ absorptions, the particular width and shape of the 2.2 $\micron$ absorption features in the spectra distinguishes hydrated silica from Al-phyllosilicates or sulfates (Figure \ref{fig:spec}B,C).
The spectra have been classified into four categories (C1-C4) based on their band depths at 1.9 and 2.2 $\micron$ to indicate different levels of spectral signature (Figure \ref{fig:A2}, Table S4).
The C1-C2 spectral classes have prominent spectral features at 1.4, 1.9, and 2.2 $\micron$ that match the band shape and width of the hydrated silica in the spectral library and lab experiments.
The C3-C4 spectra often show the broad 2.2 $\micron$ band but with a much weaker signal close to the level of noise, which does not allow a thorough examination of the shape of the absorption bands.
The presence of hydrated silica is a plausible explanation of these spectra with a 2.2-$\micron$ band, but we cannot exclude the possibility that other hydrated minerals with a 2.2-$\micron$ band, including Al-phyllosilicates, gypsum or jarosite, could be present in minor amounts in a mixture (Figure \ref{fig:spec}C).

The spectral features, including band positions and depths, \hl{can be used to identify different types of hydrated silica (e.g., opal-A, opal-CT), with implications on their origin.}
martian infrared spectra of hydrated minerals often show weaker 1.4 $\micron$ band compared to their terrestrial counterparts, resulting in small signal-to-noise ratio at this wavelength, which limits our ability to interpret the spectral band center.
However, for the C1-C2 spectral detection of \hl{silica-bearing deposits}, the 1.4 $\micron$ absorption due to Si-OH overtone occurs at $\sim 1.38 \mu m$, atypical of terrestrial laboratory spectra of opaline silica in ambient conditions of which the overtones commonly occur at 1.41 $\micron$ (Figure \ref{fig:spec}B).
In comparison with laboratory measurements of terrestrial silica of various structure and origin, the CRISM spectra with a minimum band position at 1.38-1.4 $\micron$, and the shape of the 2.2 $\micron$ feature is consistent with immature silica (e.g., hydrated glass or opal-A) under martian atmospheric condition (Figure \ref{fig:spec}\hl{DE}).
\hl{The minimum positions of 1.4 $\micron$ band used to identify different types of silica are controlled by the mineral chemistry and do not overlap with wavelength regions that are affected by artifacts and atmospheric residuals. The band shapes of 1.9 and 2.2 $\micron$ features however may be altered if there is significant atmospheric residual at 2 $\micron$ due to \ch{CO_2} compared to the absorption features. Therefore we highlight C1,C2 spectra with well-defined absorption bands so the influence by noise or atmosphere is minimal (Figure \ref{fig:spec}).}
In the meantime, the 1.4 and 1.9 $\micron$ band shapes constrained by the CRC parameters are consistent with the spectral features of silica in terrestrial sediments formed during low-temperature weathering processes (Figure \ref{fig:spec}F, Figure \ref{fig:A4}).
\hl{Having tested a large dataset (Figure \ref{fig:A4}), we highlight those with a more robust signal (Figure \ref{fig:spec}F), since the CRC parameters reflects subtle concavity change which is prone to noise or atmospheric residual.
Other hydrated minerals may also alter the band shape, resulting in small variations in CRC parameters, but their impact should be minor since their presence cannot be resolved using present orbital data.}

\begin{figure}[htb!] \label{fig:overlay}
    \centering
    \includegraphics[width=\textwidth]{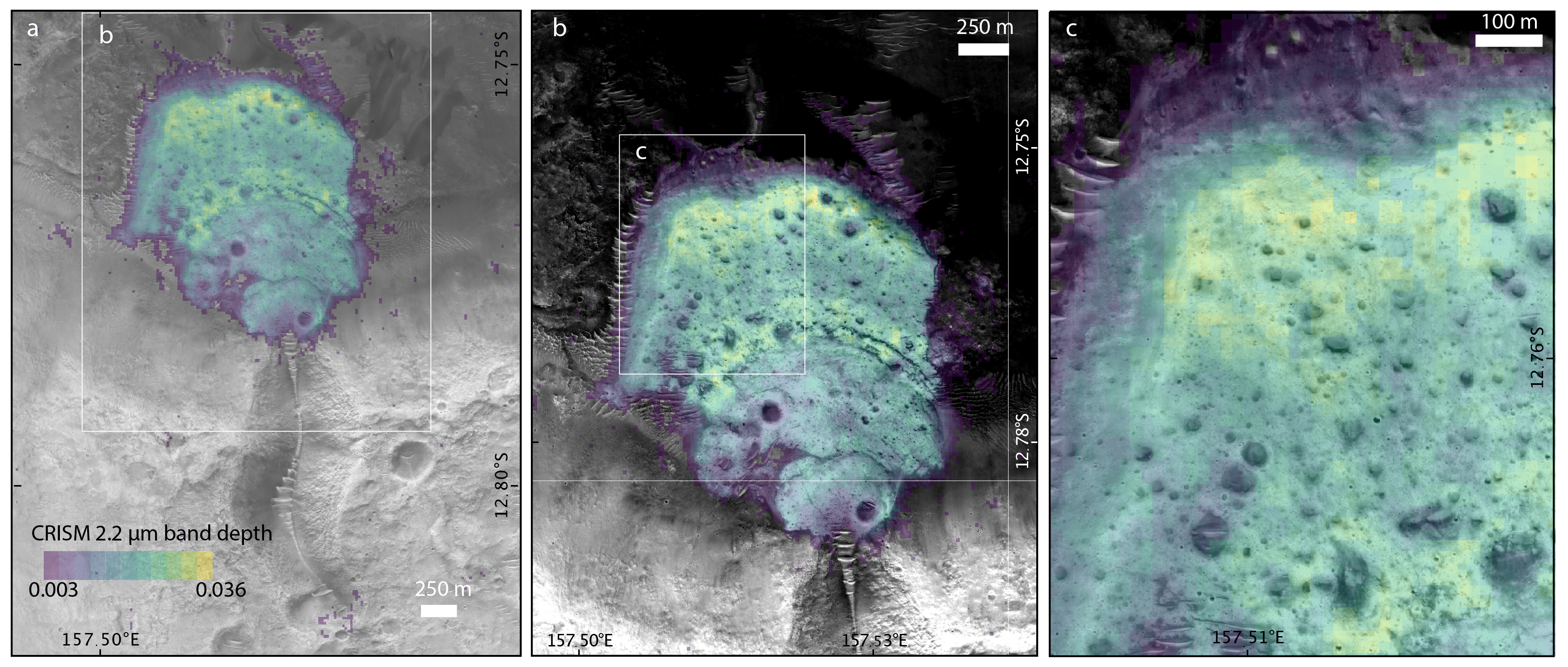}
    \hl{
    \caption{Images fans and deltas that are found with hydrated silica, with CRISM spectral parameter overlay showing the locations of the silica detections.  a. the overall context of the fan/delta including inlet channel(s); b. the fan/delta deposit with silica detection; c. a zoom image showing the silica detection. The complete figure set (6 images) is available in the online journal.} }
\end{figure}

\begin{figure}[htb!]
    \centering
    \includegraphics[width=0.95\textwidth]{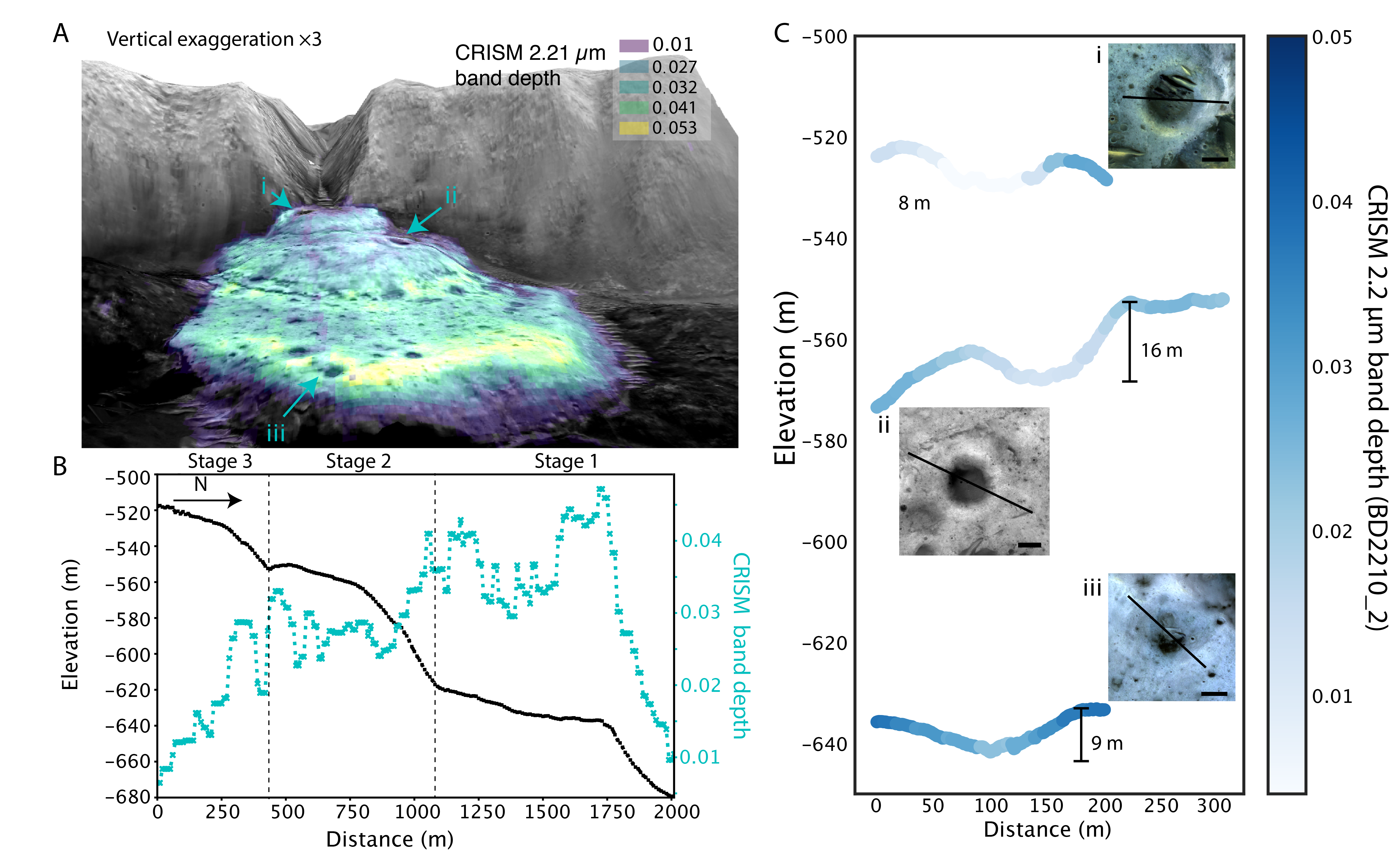}
    \caption{Elevation profile of silica detections in Aeolis fan. Strong silica detections (up to 5\%) are identified in the fan deposit. A. 3D exaggeration of Aeolis fan, overlain by the CRISM spectral parameter (BD2210\_2) indicating the band depth of 2.2 $\micron$ absorption (interpreted as Si-OH). \hl{Notations i, ii, iii refer to the craters whose cross-sectional profiles are plotted in C.} B. The central elevation profile of the same fan from apex to the toe of the crater in black and the corresponding CRISM parameter in cyan. C. Cross-sections of impact craters where walls are exposed. The floor of the craters is all filled with sand, so the variations of band depths on the crater walls are more reliable indicators of the bedrock. The silica absorption band depth decreases further into depth, but the amount of decrease is relatively small, in comparison with the changing strengths of absorption with elevation.}
    \label{fig:aeolis}
\end{figure}

\begin{figure}[htb!]
    \centering
    \includegraphics[width=0.5\textwidth]{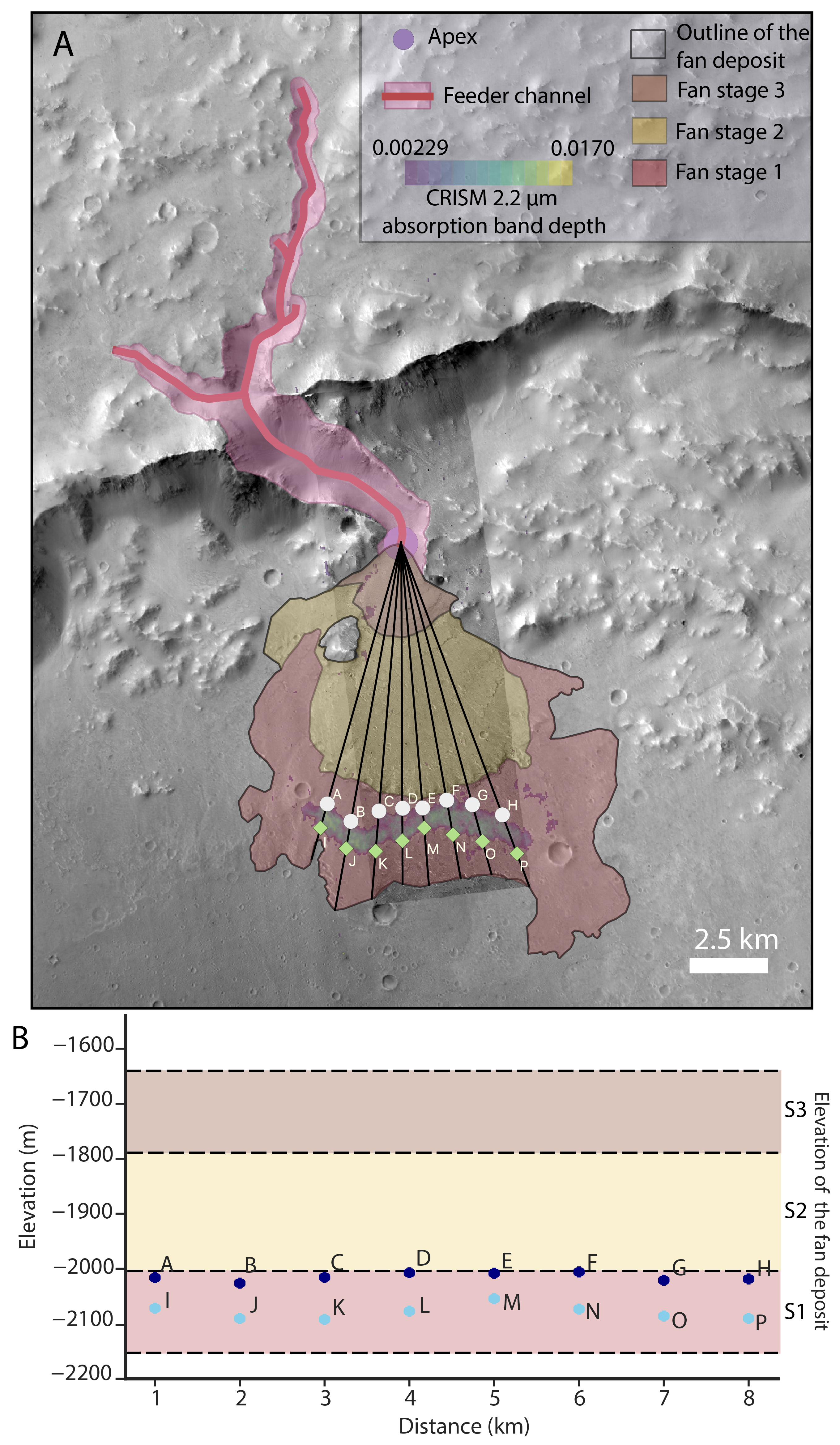}
    \caption{Elevation of the \hl{silica-bearing deposits} of the Camichel fan. A. Geological map showing the location of different fan stages and the feeder channel connected to the fan. The base map is HiRISE image PSP\_006941\_1825 and CTX mosaic. B. The elevation plot of silica detection with eight different profiles from the apex of the fan to the base of the fan. A-H are intersections with the upper level of the silica deposit, and I-P are the lower intersections. Compared to the elevation of the fan, the \hl{silica-bearing deposits} occur at constant elevation levels. }
    \label{fig:camichel}
\end{figure}

\begin{figure}[htb!]
    \centering
    \includegraphics[width=0.75\textwidth]{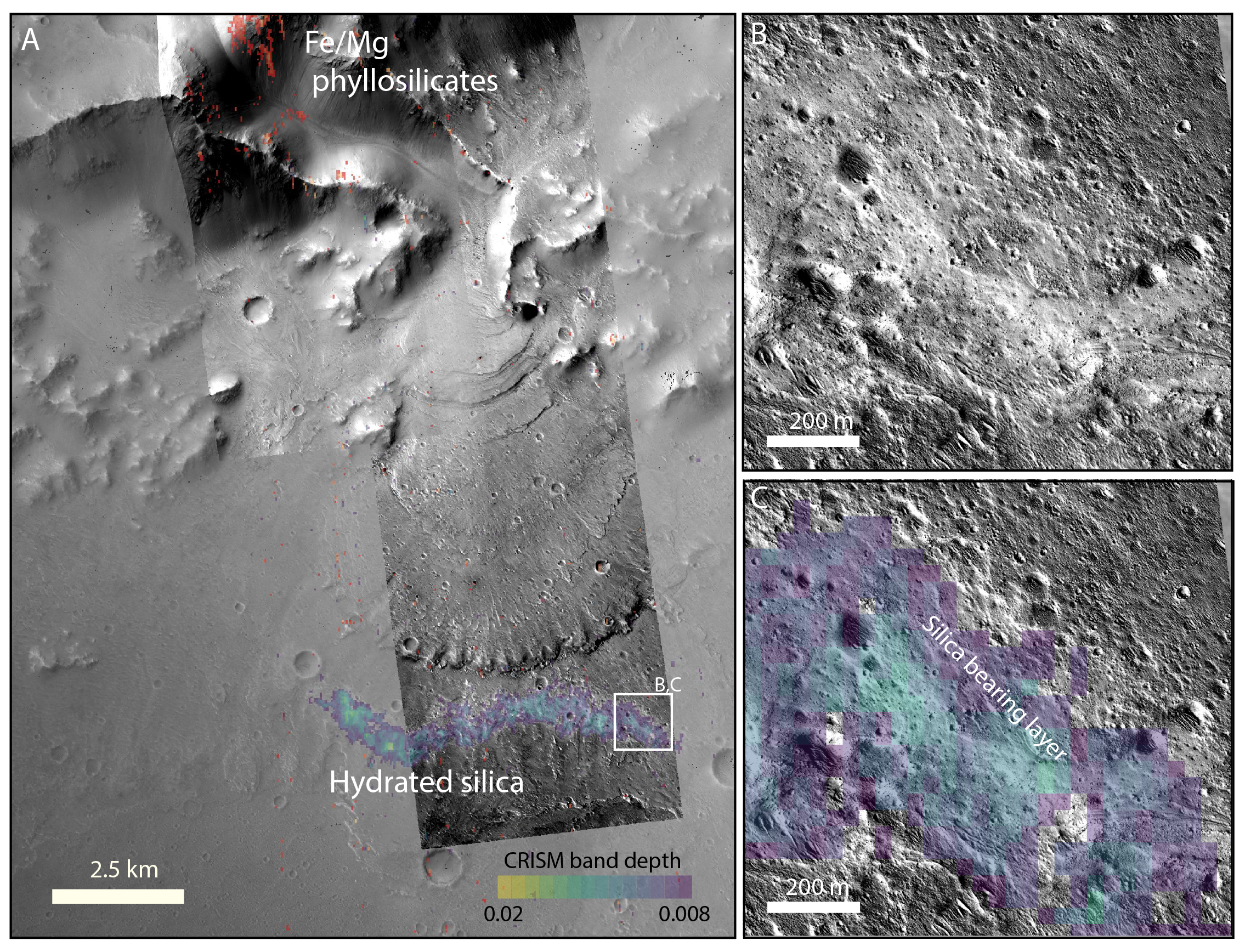}
    \caption{HiRISE images (ESP\_027287\_1830 and PSP\_006941\_1825) registered on the CTX mosaic over Camichel fan. A. The Fe/Mg phyllosilicate detections on the channel walls and silica deposit near the base of the fan. White box shows the location of B and C. B. Close-in on the HiRISE image ESP\_027287\_1830 showing the texture of the silica bearing deposit. C. The same extent as B with overlay of CRISM spectral paramters of 2.2 $\micron$ band depth showing the location of the silica detection. Note the distinct smooth, fine-grained texture and higher albedo relative to upper and lower beds. The strongest silica detection (light blue) is associated with the smooth, light-toned layer on HiRISE image.}
    \label{fig:zoom}
\end{figure}

The \hl{silica-bearing deposits} inside fan/deltas are associated with layered deposits, and their spectral signatures are correlated with elevations, as in the case of Aeolis (Figure \ref{fig:overlay}A), Camichel (Figure \ref{fig:overlay}B), and Garu (Figure \ref{fig:overlay}C).
The elevation profiles suggest the correlation of these siliceous layers with the sedimentary layers within the fans, but no visible outcrop of planar bedding is found at these locations.
Other deposits of silica occur in spatially discontinuous patches in \hl{distal sediments, and are correlated with lower elevation or greater distance to the source}, as in the case for Eberswalde (Figure \ref{fig:overlay}-4), Jezero (Figure \ref{fig:overlay}-5), and Oxia Planum (Figure \ref{fig:overlay}-6).
Aeolis, for example, is a stepped fan/delta with robust silica detection corresponding to the entire fan deposits.
There are three topographic steps in the fan, which show varying Si-OH band depth with a continuous increasing 2.21 $\micron$ absorption band depth with decreasing elevation (Figure \ref{fig:aeolis}A,B).
In Camichel and Garu fans, the \hl{silica-bearing deposits} are found in discrete, continuous horizontal layers of ~50 meters in thickness (Figure \ref{fig:overlay}-2,3, \ref{fig:camichel}), correlated with fine-grained, smooth texture in HiRISE images (Figure \ref{fig:zoom}).
Cross-sections on the walls of small impact craters expose subsurface materials, which show silica spectral signature is present along the rim down to $\sim$10 m depth (Figure \ref{fig:aeolis}C).
In Camichel and Garu fans, the silica deposits are found only in discrete, continuous horizontal layers of $\sim$50 meters in thickness (Figure 3B, C, 5), correlated with fine-grained, smooth texture in HiRISE images (Figure \ref{fig:zoom}).
At Claritas Fossae and Amazonis \hl{silica-bearing deposits} are also found over layers of sedimentary deposits, but with weaker and less continuous spectral signatures.
At these locations, silica could be exposed locally from dust cover or less well-preserved, inhibiting our ability to assess the distribution of \hl{silica-bearing deposits} in these images fully.
At the other fan/deltas, hydrated silica is only found in small regions of interest (10-50 pixels), preferentially in the distal region, outside of the bulk fluvial deposits (Table \ref{tab:1}).
In the five stepped fan/deltas where silica detections are correlated with elevation, the volume of the silica-bearing sedimentary rocks varies from 0.15 $km^3$ to 2.7 $km^3$ (Table S7).

The small surface areas and significant denudation through time prevent an accurate dating of the bulk fan deposits \citep{palucis_quantitative_2020}, but
in many cases, an estimate may be obtained based on the largest crater and stratigraphic relations.
\hl{The ages of the deltas and fans span from Noachian to Amazonian (See Table \ref{tab:1} and references therein).
It is rather intriguing that the most spectrally remarkable and spatially extended silica-bearing deposits are found in stepped fans/deltas that formed mostly in the last 3.5 Ga (Table \ref{tab:1}).
Six stepped fan/deltas in our survey are found with hydrated silica, including Aeolis, Camichel, Garu, Amazonis, Claritas Fossae, and Coprates.
These locations are found with a bulk silica-bearing deposit inside the fan/delta deposits, except for Coprates where the silica is identified in the distal region outside the bulk deposits. Other than Coprates, all these sedimentary features likely formed in the last 3.5 Ga: e.g., Camichel (0.57 Ga), Garu (0.4 Ga-3.46 Ga), Amazonis (0.68 Ga).
The tectonic features in the regional context of Claritas Fossae are dated to be Hesperian, suggesting the fluvial activity was active at or after the Hesperian \citep{mangold_detailed_2006}.
The Aeolis fan is deposited on a crater floor unit where the nine largest craters give a crater count age of 3.5 Ga with continuous resurfacing at least until 0.5 Ga (Figure \ref{fig:A7}).}

\hl{On the contrary, most of the older deltas (e.g., Eberswalde, Jezero) and fans (e.g. Murray, Holden) have silica-bearing deposits in very localized outcrops and discontinuous patches in the distal region of the delta and the silica spectral features are relatively weaker.}
Given the lack of complete outcrop of these silica-bearing deposits, we cannot exclude the possibility that the apparent differences in spatial extent and spectral signature may be related to the preservation states, rather than a primary feature of the outcrop.
If fan/delta formation involved multiple stages of fluvial activities, the crater-count based age of the fans and deltas at these locations only give an upper limit to the age of silica formation.

\hl{In addition, there are other locations including Bradbury, Majuro and possibly Maja where silica are identified in the bulk sedimentary deposit, but do not have the typical stepped morphology.
The distinct geologic context and aqueous activity at each location will have to be investigated in detail in future studies to elucidate their relationships to mineralogy. }
While these fan/deltas' formations are scattered in time, given their well-preserved features and stratigraphic relations observed, they likely indicate the last active fluvial activity in the fluvial system they are connected to \citep[e.g.,][]{mangold_fluvial_2020}.

\figsetstart
\figsetnum{3}
\figsettitle{(See Appendix B) Locations of hydrated silica detection}

\figsetgrpstart
\figsetgrpnum{3.1}
\figsetgrptitle{CRISM overlay of silica detection on Aeolis fan}
\figsetplot{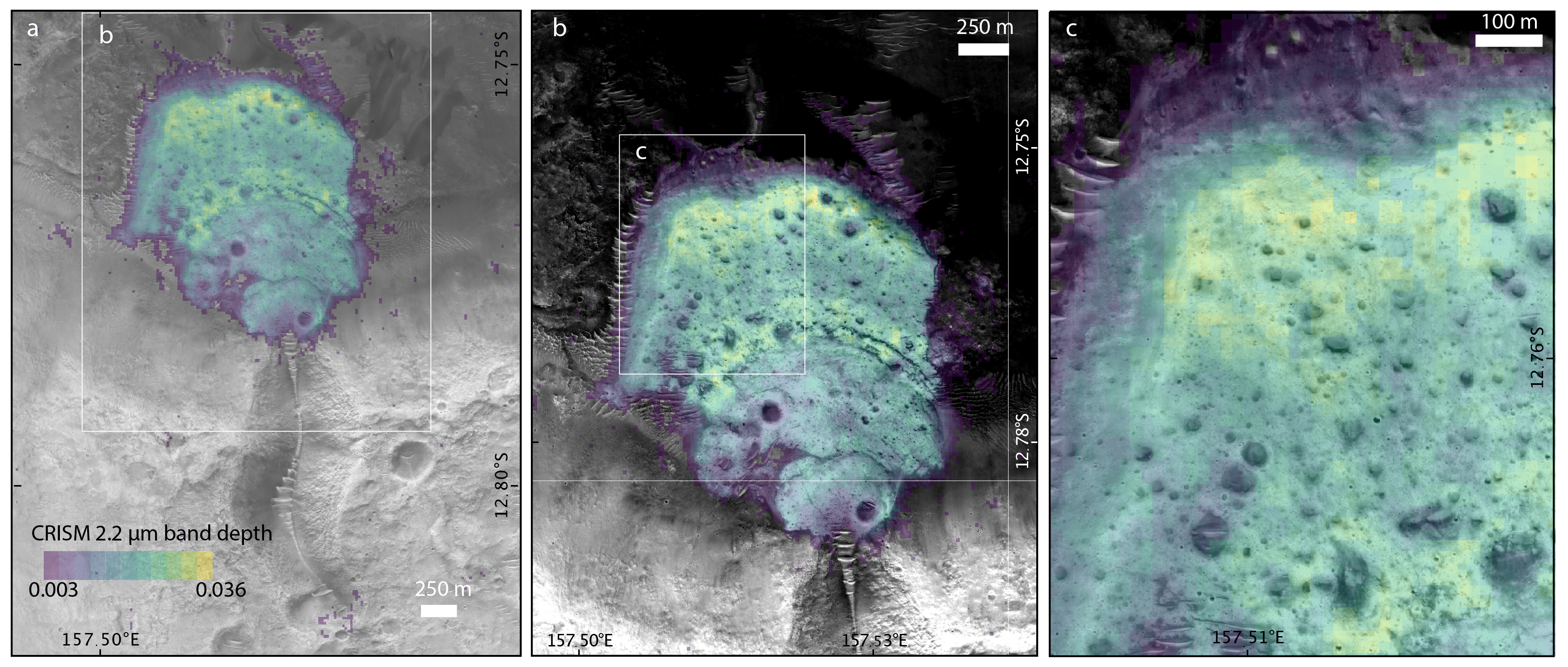}
\figsetgrpnote{Images fans and deltas that are found with hydrated silica, with CRISM spectral parameter overlay showing the locations of the silica detections. a. the overall context of the fan/delta including inlet channel(s); b. the fan/delta deposit with silica detection; c. a zoom image showing the silica detection.}
\figsetgrpend

\figsetgrpstart
\figsetgrpnum{3.2}
\figsetgrptitle{CRISM overlay of silica detection on Camichel fan}
\figsetplot{Figure3_r1-02.png}
\figsetgrpnote{Images fans and deltas that are found with hydrated silica, with CRISM spectral parameter overlay showing the locations of the silica detections. a. the overall context of the fan/delta including inlet channel(s); b. the fan/delta deposit with silica detection; c. a zoom image showing the silica detection.}
\figsetgrpend

\figsetgrpstart
\figsetgrpnum{3.3}
\figsetgrptitle{CRISM overlay of silica detection on Garu fan}
\figsetplot{Figure3_r1-03.png}
\figsetgrpnote{Images fans and deltas that are found with hydrated silica, with CRISM spectral parameter overlay showing the locations of the silica detections. a. the overall context of the fan/delta including inlet channel(s); b. the fan/delta deposit with silica detection; c. a zoom image showing the silica detection.}
\figsetgrpend

\figsetgrpstart
\figsetgrpnum{3.4}
\figsetgrptitle{Silica detection associated with Eberswalde delta}
\figsetplot{Figure3_r1-04.png}
\figsetgrpnote{Images fans and deltas that are found with hydrated silica, with CRISM spectral parameter overlay showing the locations of the silica detections. a. the overall context of the fan/delta including inlet channel(s); b. the fan/delta deposit with silica detection; c. a zoom image showing the silica detection.}
\figsetgrpend

\figsetgrpstart
\figsetgrpnum{3.5}
\figsetgrptitle{Silica detection associated with Jezero delta}
\figsetplot{Figure3_r1-05.png}
\figsetgrpnote{Images fans and deltas that are found with hydrated silica, with CRISM spectral parameter overlay showing the locations of the silica detections.  a. the overall context of the fan/delta including inlet channel(s); b. the fan/delta deposit with silica detection; c. a zoom image showing the silica detection.}
\figsetgrpend

\figsetgrpstart
\figsetgrpnum{3.6}
\figsetgrptitle{Silica detection associated with Oxia delta}
\figsetplot{Figure3_r1-06.png}
\figsetgrpnote{Images fans and deltas that are found with hydrated silica, with CRISM spectral parameter overlay showing the locations of the silica detections.  a. the overall context of the fan/delta including inlet channel(s); b. the fan/delta deposit with silica detection; c. a zoom image showing the silica detection.}
\figsetgrpend

\figsetend

\begin{deluxetable}{ccccccccccc}
\label{tab:1}
\tablewidth{\textwidth}
\tabletypesize{\scriptsize}

\tablecaption{Locations with hydrated silica identified in proximity to fluvial morphology$^{\#}$.}
\tablenum{1}
\tablehead{ \colhead{Latitude} & \colhead{Longitude} & \colhead{Name} & \colhead{Basin$^\dag$} & \colhead{CRISM ID} & \colhead{Spectra} & \colhead{Morphology} & \colhead{Location$^!$} & \colhead{ Reference} & \colhead{Reference} & \colhead{Age$^\&$} \\
\colhead{($^o$N)} &
\colhead{($^o$E)} & \colhead{} & \colhead{} & \colhead{(e.g.,)} & \colhead{(Category)} & \colhead{} & \colhead{} & \colhead{(fan/delta)} & \colhead{(silica)} & \colhead{Ga (ref.)}}
\startdata
-12.7 & 157.5 & Aeolis & C & FRT000086B7 & C1 & Stepped & Bulk &  & 1, 2 & $<3.5^*$ \\
2.7 & -51.7 & Camichel & C & HRL0000927F & C1 & Stepped & Bulk & 3-6 & 4,6 & 0.57 (4) \\
-6.6 & 141.2 & Garu & C & HRL0000C549 & C1 & Stepped & Bulk & 4,7 & 1 & 3.46-0.4 (4) \\
-39.2 & -103.1 & Claritas Fossae & C & FRT0000944A & C3 & Stepped & Bulk & 8 & 1 & $<$ H (8) \\
-8.0 & -146.6 & Amazonis & C & FRT0001BB1F & C4 & Stepped & Bulk & 9-10,6 & 1 & 0.6-0.8$^{**}$ (9) \\
-5.1 & 132.9 & Sharp & C & FRT000064CE & C4 & Fan & Distal & 11 & 1 &  \\
22.2 & 66.9 & Baldet & C & FRT00008D46 & C3 & Fan & Distal & 12 &  &  \\
-23.8 & -33.6 & Eberswalde & C & FRT00009C06 & C4 & Delta & Distal & 13-14 & 1 & 3.65-3.46 (49) \\
18.5 & 77.4 & Jezero & C & HRL000040FF & C4 & Delta & Distal & 15-17 & 1, 18 & $>3.5$ (50) \\
17.8 & -53.7 & Maja & P & HRL0000B48B & C2 & Channel & Channel &  & 1 & 3.5$^{***}$ (19)\\
-15 & -60.3 & Coprates & O & FRT00007203 & C4 & Stepped & Distal & 20-22 & 23 & $<3.16$ (22) \\
3.2 & 85.9 & Bradbury & P & FRT0000B0CB & C3 & Fan & Bulk & 24-25 & 26-28 & $<3.83$ (28) \\
17.9 & -24.0 & Oxia & P & ATU0003D04C & C2 & Delta & Distal & 29-30 & 29-30 & $>3$ (26) \\
-9.8 & -76.5 & Melas & O & FRT00009B66 & C3 & Erosional$^\S$ & Distal & 31-33 & 33-34 & $>3$ (32) \\
8.4 & -49.1 & Tyras & C & HRL0000AB77 & C3 & Delta & Distal & 35 & 1 & 3.35-3.63 (4) \\
-26.7 & -34.5 & Holden & C & FRT0000474A & C4 & Fan & Distal & 36-38 &  & 3.65-3.46 (49) \\
-22.0 & 66.7 & Harris & C &  HRL00013C12 & C3 & Fan & Distal & 38-39 &  & lN (39) \\
2.1 & 121.9 & Nepenthes & P & FRT0001D7F0 & C2 & Delta & Distal$^\ddag$ & 40-41,4 & 1 & 1.80  (4) \\
-40.3 & -175.2 & Simois Colles & P & FRT00016E5C & C2 & Channel & Channel &  & 1 &  \\
35.9 & -8.2 & Eden Patera & O & FRT0000CE47 & C1 & Other & Other &  & 1 &  \\
36.7 & -0.1 & Oxus & O & FRT0000A3D8 & C4 & Channel & Channel &  & 1 &  \\
44.7 & 8.4 & Semeykin & P & FRT00009E68 & C4 & Other & Other &  & 1 &  \\
-33.2 & 84.5 & Majuro & C & FRT0001642D & C3 & Fan & Bulk,Distal & 12 & 46 & 3.59 (42) \\
-5.4 & 137.0 & Gale & C & FRT000045F2 & C3 & Fan, delta & Other & 9,43-44 & 45-46 & $<3.6$ (43-44) \\
-28.5 & -51.3 & Ritchey & C &  FRT0000AC1F & C1 & Other & Other & 47-48 & 48 & 3.5 (47) \\
-27.17 & 73.82 & Terby & C & FRT00009A8D & C4 & Delta/Other$^\$$ & Distal & 51-52 & 52 & eN-H (52) \\
19.7 & 80.08 & Nili & P & FRT00016655 & C3 & Channel & Other & & & \\
-23.62 & 27.98 & Murray & C & FRS0003D1A2 & C3 & Fan & Distal & & & \\
\enddata

\tablecomments{
$^\#$ Locations where multiple fans and deltas are found (e.g., Harris, Melas) have been grouped together in this table.\\
$^\dag$ Abbreviation for basin type: Crater (C), Plains (P), and Others (O).\\
$^\S$ Erosional remnant of a fan-shaped deposit, where only the top surface is exposed.\\
$^!$ Silica location is categorized as ``Fan", ``Distal", ``Channel" or ``Other". \hl{``Bulk"} indicates silica detection on the bulk deposit of the fan/delta, while ``Distal" indicates silica detected outside of the bulk fan deposit, in the distal regions of the fan/delta. ``Channel" suggests silica is detected within the fluvial channel. ``Other" indicates silica detection in close vicinity of a fluvial fan or delta, but they are not directly related to the fluvial morphology, including occurrences on the central peak or central mound of a crater, layered deposits, or the bedrock close to fluvial features.\\
\hl{$^\ddag$ Nepenthes is a Gilbert-type delta on top of older deltaic deposits. Silica outcrop at distal locations may belong to older deposits where the fluvial morphology is unclear.\\}
$^\$$ Terby crater hosts a thick sequence of light-toned layered deposits but lacks the fluvial channels leading to the deposits, which have been interpreted as deltaic deposits \citep{ansan_stratigraphy_2011} or volcanic/aeolian deposits \citep{wilson_geomorphic_2007}. \\
$^\&$ Age of the fan or delta deposits from previous literature. Abbreviations: H-Hesperian; lN-Late Noachian.\\
$^*$ \hl{Age of the floor unit.} Crater count over the fan deposits is saturated.\\
$^{**}$ Age of the lake basin based on crater density of the basin floor.\\
$^{***}$ Age of the Maja terminus and Chryse crater floor while younger resurfacing events have occurred \citep[e.g.,][]{chapman_possible_2003}.
}

\tablerefs{1. \citet{carter_composition_2012}; 2. \citet{sun_distinct_2018}; 3. \citet{hauber_sedimentary_2009}; 4. \citet{hauber_asynchronous_2013}; 5. \citet{popa_new_2010}; 6. \citet{goudge_constraints_2012}; 7. \citet{di_achille_ancient_2010}; 8. \citet{mangold_detailed_2006}; 9. \citet{cabrol_evolution_2001}; 10. \citet{hughes_evaluation_2012}; 11. \citet{irwin_sedimentary_2004}; 12. \citet{kraal_catalogue_2008}; 13. \citet{malin_evidence_2003}; 14. \citet{moore_martian_2003}; 15. \citet{fassett_fluvial_2005}; 16. \citet{tarnas_orbital_2019}; 17. \citet{masursky_classification_1977}; 18. \citet{baker_martian_1979}; 19. \citet{chapman_possible_2003} ;20. \citet{weitz_formation_2006}; 21. \citet{di_achille_steep_2006}; 22. \citet{grindrod_formation_2012}; 23. \citet{weitz_mixtures_2015}; 24. \citet{erkeling_valleys_2012}; 25. \citet{bramble_testing_2019}; 26. \citet{bishop_phyllosilicate_2008}; 27. \citet{bishop_mineralogy_2013}; 28. \citet{tirsch_geology_2018}; 29. \citet{quantin-nataf_exomars_2019}; 30. \citet{carter_oxia_2016}; 31. \citet{weitz_geology_2003}; 32. \citet{quantin_fluvial_2005}; 33. \citet{metz_sublacustrine_2009}; 34.  \citet{williams_reconstructing_2014}; 35. \citet{di_achille_geological_2006}; 36. \citet{grant_drainage_2002}; 37. \citet{pondrelli_complex_2005}; 38. \citet{moore_large_2005}; 39. \citet{williams_evidence_2011}; 40. \citet{irwin_intense_2005}; 41. \citet{kleinhans_palaeoflow_2010}; 42. \citet{mangold_late_2012}; 43. \citet{thomson_constraints_2011}; 44. \citet{le_deit_sequence_2013}; 45. \citet{seelos_mineralogy_2014}; 46. \citet{fraeman_stratigraphy_2016}; 47. \citet{ding_central_2015}; 48. \citet{sun_geology_2014}; 49. \citet{mangold_origin_2012}; 50. \citet{mangold_fluvial_2020}; 51. \citet{wilson_geomorphic_2007} ; 52. \citet{ansan_stratigraphy_2011}.}

\end{deluxetable}

\section{The origin of hydrated silica in fans and deltas}\label{sec:discuss}
Hydrated minerals, mostly Fe/Mg phyllosilicates, have been previously identified within fans and deltas, where proposed origins for these hydrated minerals have been put forward as either detrital \citep[e.g.,][]{ehlmann_clay_2008,murchie_synthesis_2009,milliken_sources_2010,poulet_mineral_2014} or authigenic \citep[e.g.,][]{dehouck_ismenius_2010,bristow_terrestrial_2011,ansan_stratigraphy_2011,mangold_late_2012,hauber_asynchronous_2013,poulet_mineral_2014}.
Unlike Fe/Mg phyllosilicates, hydrated \hl{silica-bearing deposits} may also occur in the form of volcanic glass, hydrothermal deposits (including sinters), as in the case for silica detections in volcanic calderas in Nili Patera \citep{skok_silica_2010}.
The spectral signature, the spatial distribution, and the volumetric measurements in comparison with sedimentary deposits are key to unravel the geologic process in which \hl{silica-bearing deposits} were formed in martian alluvial fans and deltas.

\subsection{The spectral features of hydrated silica}
The shorter band center of the 1.4 $\micron$ absorption feature in association with the broad 2.2 $\micron$ band matches spectral features with deposits enriched in amorphous or dehydrated silica.
Possible explanations of the spectral signature include the formation of hydrated glass \citep{rice_reflectance_2013,smith_hydrated_2013}, as well as a dehydrated opal-A under martian atmospheric conditions \citep{sun_distinct_2018,poitras_mars_2018} or elevated temperature \citep{bobon_state_2011}.
Here the geologic context (co-occurrence with fluvial deposits and smooth, high albedo features) supports the origin as a silica deposit from water-rock interaction rather than detrital glass.
The susceptibility to weathering of volcanic glass also means they are less likely to persist in a prolonged fluvial setting.
Additionally, the shape of the 1.4 and 1.9 $\micron$ characterized by the CRC parameter shows that these CRISM spectra are more consistent with terrestrial silica (opal A/CT) from low-temperature weathering \citep{pineau_toward_2020}, for select locations where 1.4 and 1.9 $\micron$ bands are well-defined and not affected by atmospheric residual.
The spectral signature including the shorter wavelength position of 1.4 $\micron$ band, the lack of concavity of the 1.9 $\micron$  and 2.2 $\micron$  bands agree with that of relatively pristine and dehydrated opal-A, as compared to the silica spectra that are often found in eolian/transported deposits on Mars, consistent with opal-CT \citep{sun_distinct_2018,pineau_toward_2020}.
The spectral characteristics support the interpretation of dehydrated opal-A formed in low-temperature weathering processes ($<50^oC$), congruous with the geologic context of altered fluvial deposits.

\hl{
Such an immature spectral feature indicate the hydrated silica associated with these fans and deltas may be primary deposits, in contrast to the mobilized silica in present-day eolian sediments as shown by \citet{sun_distinct_2018}.
Opal-A bearing deposits on Earth transform to opal-C within a few million years at temperature of $50^oC$ \citep{compton_porosity_1991}.
Albeit the rate constant of such transition at low temperature is rather uncertain, a complete opal-CT to quartz conversion should occur within 100-400 Ma on martian conditions given reasonable assumptions \citep{tosca_juvenile_2009}.
The presence of such immature type of silica indicated by the spectral features associated with fans or deltas thus imply very limited duration of water availability, if at all, after their initial deposition.
}

\subsection{Distribution and context of silica in association with fluvial morphology}
The 2.2-$\micron$ band depth indicative of hydrated silica is found to be correlated with the elevation of the deposit in Aeolis fan (Figure \ref{fig:model}A, \ref{fig:aeolis}) and formed as a continuous layer in Camichel and Garu (Figure \ref{fig:overlay}B,C), in concordance to the layering when observed.
The variations in band depth could be due to changes in lighting geometry, exposures of fresh surfaces from dust, as well as hydrated silica concentration and grain size.
Lighting geometry influences the reflectance level observed, but it is found that the spectral band depth, calculated from the ratio of reflectance, should vary only by a few percent when the phase angle is smaller than 60 degrees \citep{shepard_laboratory_2011,ruesch_near_2015}.
At Aeolis, the phase angle between 39.5-45.5 degrees could not have caused significant variations in spectral band depth. If the band depth variations are caused by dust cover, it is expected to have dust accumulation and therefore reduced band depths at lower elevation and locations with gentler slope, contrary to our observations.
Therefore, the spectral band depth variation at Aeolis is likely correlated to the physical or chemical changes in the sediments, possibly due to variations in silica concentration or grain size \citep[e.g.,][]{clark_high_1990}.
Either silica concentration increases with lower elevation, or silica is found in smaller grain-sized particles at lower parts of the fan. The distribution of silica preferentially in the distal regions is in contrast to the expected behavior at a hydrothermal source.
As temperature decreases exponentially with time and distance to the source, silica precipitation should be expected to decrease over time and with greater distance if hydrothermal fluid were flowing on the surface, without additional heating sources (Figure \ref{fig:A5}).
The thickness of the \hl{silica-bearing deposits} can be indicative of their formation mechanisms.
At Aeolis, the silica band depth decreases with depth within the craters but remains above the detection limit down \hl{to $\sim$10 m in elevation} (Figure \ref{fig:aeolis}).
Formation as a continuous layer with correlating texture and constant elevation suggests silica precipitation occurred within the ~50 meters-thick deposits in Camichel and Garu (Figure \ref{fig:overlay}, \ref{fig:camichel}, \ref{fig:zoom}).
In addition, thin coatings and rinds of 100s of microns thick would easily be removed \hl{within a million years even with the limited erosional rates (0.01-0.1 nm/yr) of Amazonian Mars \citep{golombek_erosion_2000}.}
These observations indicate that the silica detection is not limited to the upper micron-thick layer on top of these deposits, in contrast to the micron-sized silica coatings of Hawaiian basalt, which forms rather quickly as shown in previous studies \citep{minitti_morphology_2007,chemtob_silica_2010}.

The source regions of these fans and deltas \hl{lack silica-bearing outcrops}.
The lack of observation may be due to dust cover, as tens of microns of dust cover could inhibit our ability to detect hydrous minerals using CRISM data.
However, unlike Fe/Mg phyllosilicates, which are widespread in the southern highlands of Mars, \hl{silica-bearing deposits} have only been found in small, localized deposits \citep[e.g.,][]{skok_silica_2010,bishop_phyllosilicate_2008}, most of which with \hl{smaller} spatial extent than the deposits in fans and deltas.
We have not yet identified any location in which widely distributed \hl{silica-bearing deposits} occur in the source region of the fan or delta deposit, although some isolated regions have been found with hydrated silica outside the sedimentary basin (e.g., Melas and Coprates  \citep[e.g.,][]{milliken_sources_2010,le_deit_extensive_2012,carter_composition_2012}).
\hl{Unless there exists widespread silica-bearing deposits in the subsurface that has not been detected so far, the silica observed here, in particular in the stepped fan/deltas where bulk silica-bearing materials are found, likely did not inherit from an earlier deposit. }

\subsection{\hl{The origin of silica-bearing deposits}}
\hl{As discussed in the previous sections, there is no identified widespread silica deposits indicating the hydrated silica-bearing deposits could have been transported from elsewhere as detrital sediments.
If the silica deposits were buried in the subsurface within the source rocks, they might have acted as a lubricator resulting in the basal slide similar to the effect of clay minerals \citep{watkins_long-runout_2015}, thus forming fan-shaped deposits in front of a wall scarp.
The fans and deltas investigated in this study are connected to well-defined inlet channels (e.g., Figure \ref{fig:overlay}), in contrast to the typical landforms of landslides with an arcuate breakaway zone in the source region.
Thus the morphology and lack of source region for these silica-bearing deposits are inconsistent with the landslide scenario formed with subsurface silica deposits.}

\hl{Alternatively, }silica which is easily mobile during water-rock interaction could dissolve in pore-waters and re-precipitate \citep[e.g.,][]{sjoberg_silica_1996,williams_silica_1985-1,mclennan_sedimentary_2003}.
While terrestrial alluvial fans in semi-arid locations can be carbonate-cemented \citep{blissenbach_geology_1964,nickel_carbonates_1985}, different pH and water chemistry may have been involved in silica formation in alluvial fans/deltas on Mars.
The silica-bearing layers occurring at a near-constant elevation relative to current geoid at each location indicate formation directly from surface standing bodies of water or the intersection of the groundwater with the surface.
In either scenario, the precipitation of silica co-occurred with the deposition of the sediments in the alluvial fans or deltas.
\hl{Late diagenetic event could have also precipitated silica deposits, as multiple-stage diagenesis has been observed in Gale crater \citep[e.g.,][]{frydenvang_diagenetic_2017,rapin_situ_2018} and Meridiani Planum \citep[e.g.,][]{mclennan_provenance_2005,christensen_mineralogy_2004}.
To be consistent with the observed distribution of silica-bearing deposits, a pre-exisitng layer of sediments with higher porosity or permeability should be present that allows later diagenetic events to occur in a constricted sedimentary layer.
If such late diagentic events did occur, we would expect silica in locations other than the fan or delta deposits (e.g. in the floor of the basin), which has not been identified so far.
It is also unlikely a late diagenetic event would significantly change the grainsize and texture of the sedimentary layer, as shown to be distinct from sediments above and below (Figure \ref{fig:zoom}).
While we do not favor a late-diagenesis scenario for the bulk silica deposits in stepped fan/deltas, it is likely such events may have occurred in many locations investigated here where silica is present as discontinuous outcrop in the distal region.
For example, aqueous activity involve multiple lake levels in Gale crater \citep{palucis_sequence_2016}, and the hydrated silica could have formed during the latest stage of diagenesis, as shown with high-silica fracture-associated halos in the Murray formation \citep[e.g.,][]{frydenvang_diagenetic_2017}.}

\begin{figure}[htb!]
    \centering
    \includegraphics[width=0.95\textwidth]{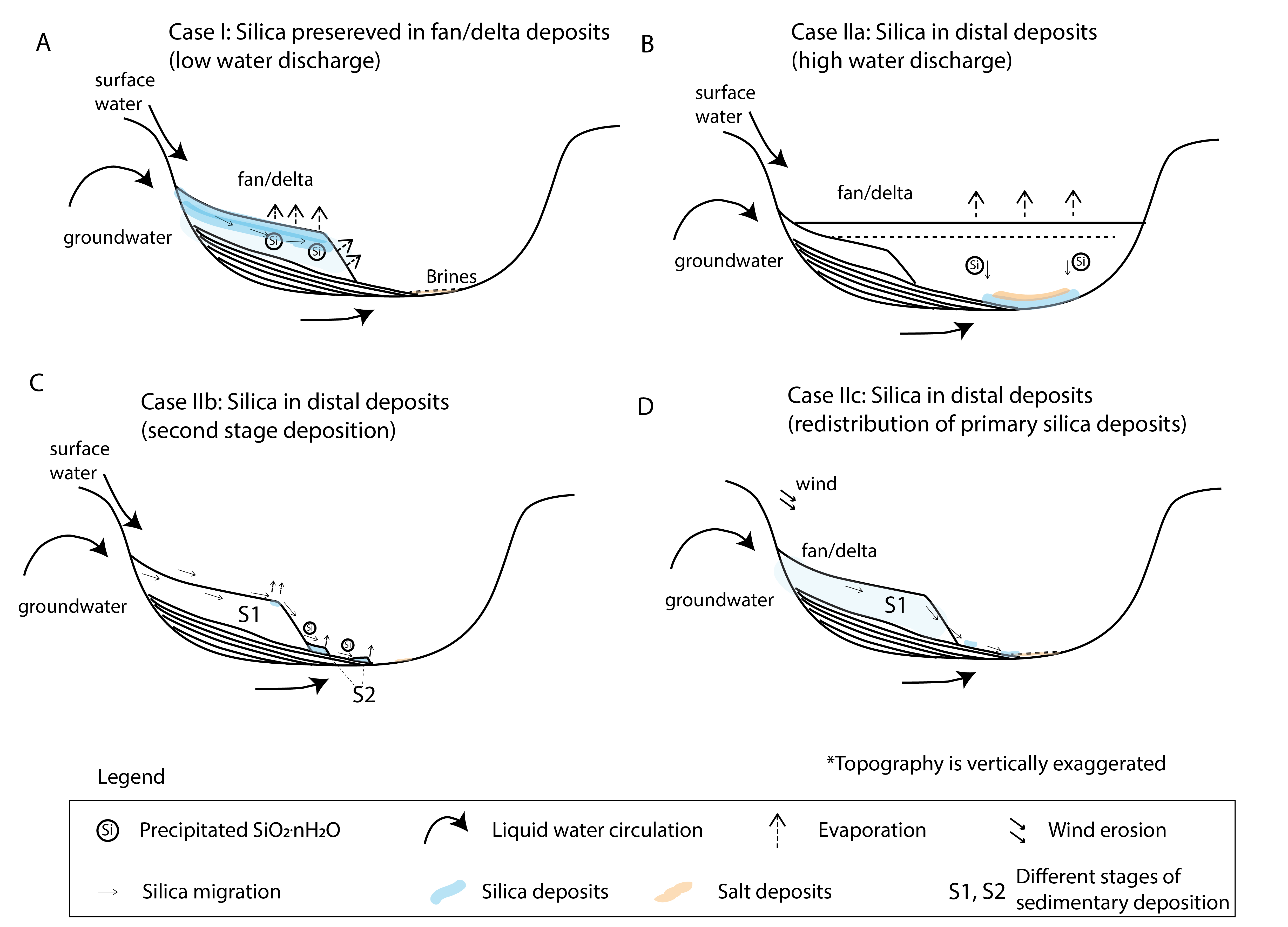}
    \caption{Schematic model of silica precipitation within fan/delta with different amounts of water discharge. The topographic profiles are vertically exaggerated to show the flow within the fan/delta deposit.}
    \label{fig:cartoon}
\end{figure}

Considering the lack of source region, the spectral signature and the correlation with topography, the formation of \hl{silica-bearing deposits likely} have occurred at the same time or shortly after the deposition of the sediments.
In particular, we propose silica precipitation in the stepped fan/deltas (e.g., Aeolis, Camichel, Garu) was active in conditions with a relatively low water discharge in the most recent epoch on Mars (Figure \ref{fig:cartoon}\textbf{A}).
In such an environment, water reaches silica oversaturation due to evaporation or change in chemical or physical conditions and begins to precipitate silica enveloping sedimentary grains within sedimentary deposits as siliceous cement.
Silica could be then accumulated as they are adsorbed to sedimentary particles and transported down section within the fan/delta deposits.
The formation of silica thus follows either the surface water upon the fan deposits or groundwater seepage at a specific level, making continuous deposits within the fan/delta.

Apart from the occurrence in the bulk sedimentary deposits in stepped fan/deltas, silica preferentially occurs at distal locations in areally-restricted, discontinuous outcrops in most older fans and deltas.
The preferred distal occurrence suggests a different condition during sedimentation or higher degrees of subsequent modifications compared to the bulk silica-bearing deposits.
During the primary fluvial process, if the water discharge is relatively high to the sediment flux and silica concentration is beneath saturation, silica precipitation within the sedimentary deposits can be hindered.
In this scenario, silica would be dissolved in the standing bodies of water, and only begin to precipitate in mixture with other salts or clay minerals on the basin floor as water evaporates (Figure \ref{fig:cartoon}\hl{B}).
\hl{Alternatively, the reactivation of a fluvial system over time \citep{hauber_asynchronous_2013,mangold_fluvial_2020} may have induced multiple sedimentary stages, and volumetrically smaller deposits may have occured after the bulk of the fan or delta have been formed, leaving localized silica deposits (Figure \ref{fig:cartoon} C).}
\hl{Otherwise, }if hydrated silica initially deposited in the bulk sedimentary deposits, they may be subject to secondary processes that redistributed silica due to a longer duration of groundwater activity, aeolian reworking and re-deposition, thus making it more challenging to preserve \hl{silica-bearing deposits} within the fan or delta (Figure \ref{fig:cartoon}\hl{D}).
\hl{In all the proposed scenarios, the silica-bearing deposits recorded the water volume and conditions during its primary precipitation, but in the older deltas or fans (Figure \ref{fig:cartoon} B-D) the current outcrops may be partially or unrelated to the primary silica deposits. Detailed context analysis \textit{in situ}, may help distinguish these scenarios which cannot be determined from orbit.}

\subsection{Mass balance with water volume estimates}
In a scenario where hydrated silica has precipitated during the fluvial activity forming the fans and deltas in place, one could make reasonable estimates of the water volume and condition based on mass balance. The volume of water involved should match between the mineralization of \hl{silica-bearing deposits} ($V_{H_2 O\_Si}$) and the formation of the geomorphologic features ($V_{H_2 O\_sed}$).

\begin{equation}
     V_{H_2 O\_sed}\sim V_{H_2 O\_Si}
     \label{eq:vsi}
\end{equation}

The first approach consists in estimating water volume through the volume of the sediment, regardless of composition. Since it is impossible to obtain a grain size distribution from orbital images, we rely on the sediment-to-water ratio ($R_{sed/water}$) to understand the order of magnitude of the volume of water involved (Eq. \ref{eq:4}).
\begin{equation}
    V_{H_2 O\_sed} = \frac{V_{sed}}{R_{sed/water}}
    \label{eq:4}
\end{equation}
Terrestrial rivers have an average sediment-to-water ratio of 1:5000 \citep{dietrich_geomorphic_2003}, while in arid or semiarid regions (e.g., Chile), the ratio is close to 1:3000 \citep{pepin_erosion_2010,palucis_origin_2014}.
However, it has been shown in the previous laboratory experiments that the typical stepped morphology observed in many martian alluvial fans (Fig. \ref{fig:overlay}-\ref{fig:zoom}) could be formed during a concentration flow with a high sediment-to-water ratio (i.e., ca. 1:10 - 1:100 \citep{kraal_martian_2008,kleinhans_palaeoflow_2010}).
As we consider the integrated volume of water throughout active fluvial processes, the sediment-to-water ratio for one specific mass flow would underestimate the integrated water volume, especially if standing bodies of water were present for a sufficiently long time.
It is also unclear if the sedimentation did occur with a very high mass concentration (ca. 1:10). Therefore, here we adopt the 1:100 ratio as a reasonable upper bound.
The uncertainties regarding the sedimentation processes indicate significant variations in the order of magnitude of the water volume involved in the formation of these fluvial deposits.
For a maximum volume of 0.15 $km^3$ of sediments that have been deposited in Aeolis, this would indicate $\sim 15$ $km^3$  (1:100) to $\sim 750$ $km^3$  (1:5000) liquid water involved in this sedimentary process.

On the other hand, the amount of silica precipitates also helps constrain the amount of water involved, where we need to consider the solubility of amorphous silica under different temperature and pH ($K_{Si} (T,   pH)$) and the concentration of silica in the sediments ($f_{Si}$) (Eq. \ref{eq:5}).
\begin{equation}
    V_{H_2O\_Si} = \frac{V_{si} \times f_{Si}}{K_{Si}(T,pH)}
    \label{eq:5}
\end{equation}
Silica remains as monomeric silicic acid in solution at acidic to neutral pH, where the solubility for different silica species is correlated with temperature \citep{gunnarsson_amorphous_2000} (Eq. \ref{eq:6}).
\begin{equation}
    Am. Silica: \log(K) = -8.476-485.24/T-2.268*10^{-6}T^2 + 3.068 \log(T)
    \label{eq:6}
\end{equation}
Near $pH=9$, the first dissociation of silicic acid ($H_4SiO_4$) became dominant until $pH=12$ \citep{dove_chapter_1995,sjoberg_silica_1996}, following the equilibrium constant (Eq. \ref{eq:7}).
\begin{equation}
    pH + \log(\frac{[H_3SiO_4^-]}{[H_4SiO_4]}) = pK_{1,1} = 9.8
    \label{eq:7}
\end{equation}

Complications of cation species in solution forming complexes, adsorption, surface area, and particle size could control the precipitation rate of \hl{silica-bearing deposits} kinetically \citep[e.g.,][]{williams_silica_1985,williams_silica_1985-1,dove_chapter_1995}.
Here we performed our calculations considering thermodynamic equilibrium during silica precipitation process (Eq. \ref{eq:6}, \ref{eq:7}, Figure \ref{fig:A6}) without the complication of the dissolution rate and kinetics of silica precipitation.

\begin{figure}[htb!]
    \centering
    \includegraphics[width=0.75\textwidth]{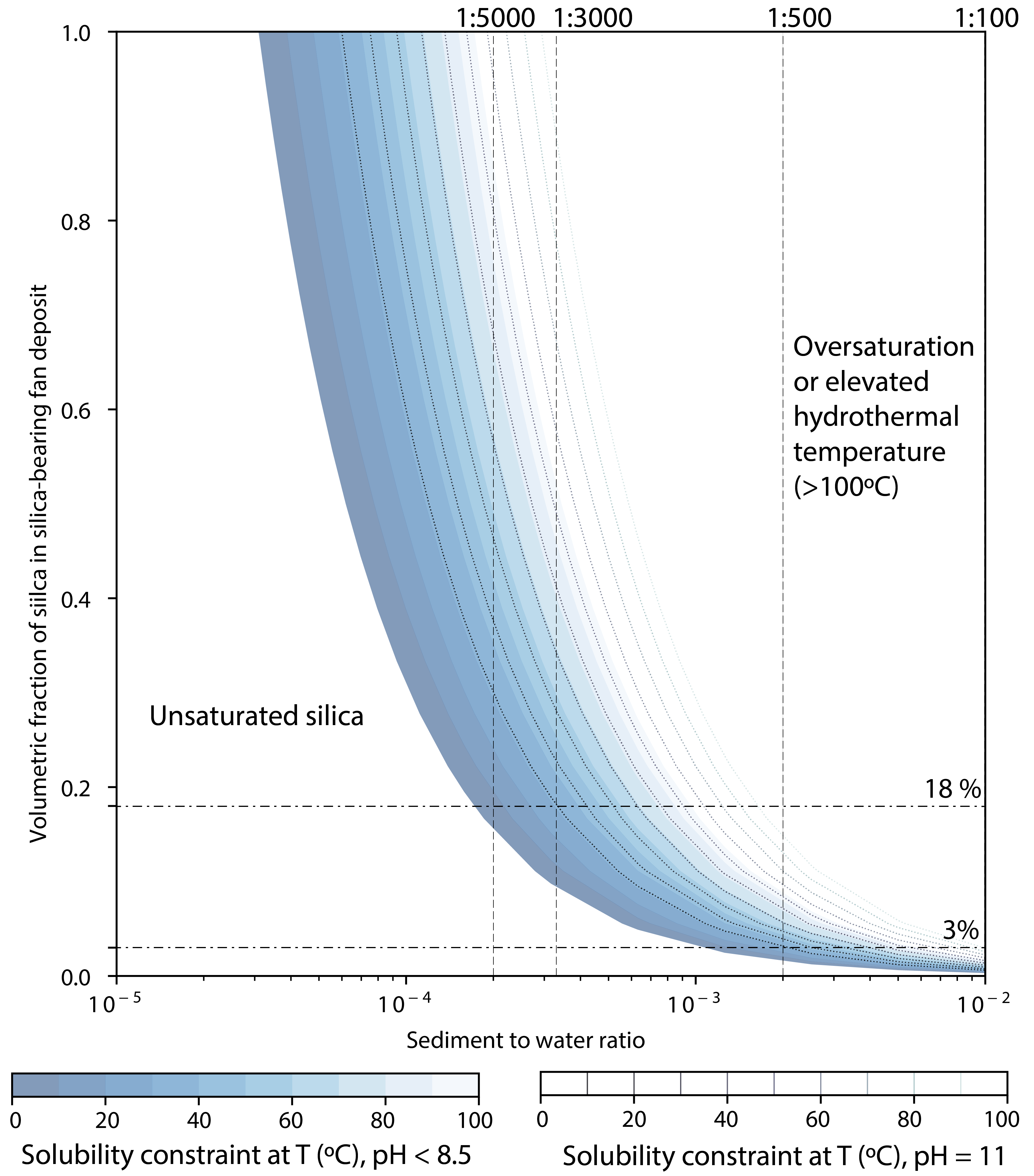}
    \caption{Estimated solubility of silica with varying volumetric fraction of silica and sediment to water ratio. The contours give the theoretical solubility constraints at temperature from 0 to 100 $^oC$ and acidic-neutral pH ($<$8.5) and pH=11 \citep{dove_chapter_1995,sjoberg_silica_1996,gunnarsson_amorphous_2000}. We show a wide range of sediment to water ratio from typical terrestrial rivers (1:5000, 1:3000) to mass concentration flows (1:500, 1:100). The range of silica solubility constraint are calculated from equilibrium constant of amorphous silica.}
    \label{fig:model}
\end{figure}

We take Aeolis as an example since all three fan steps show evidence of hydrated silica.
Given the average \hl{thickness} of the silica deposit up to $\sim 10m$ in each step of the fan, it is estimated that at least 20\% of the bulk fan comprises silica, but their concentration cannot be uniquely constrained from orbit.
We consider a reasonable range of total silica concentration within the bulk of the fan between 3\% (assuming 15\% silica in the upper 10 meters) and 18\% (assuming the upper 10 meters consist of up to 90\% silica based on observations of terrestrial siliceous sediments \citep{ledevin_les_2013,trower_sedimentology_2016}, as well as rover analysis and remote sensing \citep{squyres_detection_2008,bandfield_extensive_2013}) (Figure \ref{fig:model}).
Provided the constraints given by volumetric measurements and solubility calculations (Eq. \ref{eq:4}-\ref{eq:7}), we find silica fraction close to 18\% is consistent with a terrestrial river-like sediment-to-water ratio, while silica fraction close to 3\% may indicate a much higher sediment-to-water ratio, and therefore a debris-flow-like process of sedimentation (Figure \ref{fig:model}).
The solubility constraints do not preclude precipitation at a lower temperature ($<50 ^oC$) in any of the scenarios, consistent with the indications from CRC parameters.

\subsection{Implications for future explorations}
To refine water volume estimates, both the sediment-to-water ratio and the volumetric fraction of silica could be more precisely determined with \textit{in situ} observations that consist of high-resolution imaging and compositional analysis that provide information on the grain size distribution of the sediments and silica concentration.
\hl{Accessory minerals that might be present in minor amount would provide additional constraints on pH of water during silica formation. }
Such information, as well as detailed modeling of silica precipitation processes, would be crucial to pinpoint the environment in which these fluvial deposits and hydrated silica formed, providing novel insights into the \hl{temperature of martian waters} at the time of formation.
Other than its potential to constrain the aqueous environments, hydrated silica also has important astrobiological implications \citep{mcmahon_field_2018}.
It’s worth to note that pervasive siliceous sedimentary deposits exist in the Archean geologic records on Earth \citep{siever_silica_1962}, where textural information at the centimeter to millimeter scale and detailed compositional and isotopic analysis have been carried out to constrain the water chemistry and the formation pathways of these sediments \citep[e.g.,][]{van_den_boorn_dual_2007,ledevin_les_2013,trower_sedimentology_2016,kleine_silicon_2018}.
Comparative studies of silica samples considering the sedimentary cycles on Earth and Mars, as well as their potential to preserve organic matter in \hl{silica-bearing deposits}, make compelling goals for future analysis on returned samples from Mars.

\section{Conclusions and implications for future exploration}
As presented in this study, we identified a global process where silica precipitation occurs \hl{in close proximity with alluvial fans and deltas.}
\hl{Notably, five stepped fan/deltas which formed during late-stage fluvial activity (0.4-3.5 Ga) show evidence of silica formation concurrently with the formation of sedimentary deposits, while other older fans and deltas are identified with hydrated silica mostly in the distal region with a few exceptions.}
The hydrated silica found at these locations typically has a 1.38-1.4 $\micron$ band with broad 1.9 and 2.2 $\micron$ absorptions, consistent with amorphous silica (glass) or dehydrated opal-A.
CRC parameters indicate spectral proximity to terrestrial \hl{silica-bearing deposits} formed under low-temperature weathering conditions ($< 50 ^oC$ as defined in \citet{chauvire_near_2017}).
In a few typical locations, these \hl{silica-bearing deposits} are correlated with different sedimentary stages within the fan, follow near-planar layers and lack an obvious source.
We suggest the hydrated \hl{silica-bearing deposits} formed in stepped fans/deltas are likely authigenic products through precipitation from martian waters in close relation with the fluvial activity that resulted in the formation of fans and deltas as \hl{terminal fluvial deposits} of river channels.
The order-of-magnitude volume estimate based on the current understanding of fluvial systems on Mars is consistent with the volume of water needed to precipitate silica deposit, given a sediment-to-water ratio between 1:100 and 1:5000 during the aqueous episode forming the sedimentary deposits.
In the case of Aeolis, for example, it is possible to precipitate the required volume of silica in a relatively low temperature environment, consistent with the spectral parameter calculations.
\hl{The immature nature of these hydrated silica identified suggest very limited timescale (e.g., a few million years) for these aqueous activity that resulted in opaline silica formation and they are likely the latest aqueous events at each location.}
Given new measurements of grain size distributions and silica concentration in future in situ observations, it would be possible to constrain the volume, temperature and chemistry of the water that formed these silica-bearing fan and delta deposits.
The study highlights the potential of these \hl{silica-bearing deposits} to understand past climate conditions on Mars, in addition to their astrobiological potentials to incorporate and preserve organic molecules, demonstrating their necessity to be considered as critical scientific targets for the future Mars missions including rover explorations and Mars Sample Return.

\acknowledgments
The authors would like to thank the instrument teams for CRISM, HiRISE and CTX for providing the valuable datasets and Jay Dickson and the Bruce Murray Lab at Caltech for generating and sharing the CTX global mosaic. This project has received funding from the European Union’s Horizon 2020 research and innovation program under the Marie Sklodowska Curie grants agreement No. 751164 and ANR project ANR-18-ERC1-000.

\bibliography{sample63}{}
\bibliographystyle{aasjournal}

\appendix
\restartappendixnumbering
\section{Supplementary figures: Figure A1-A7}
\begin{figure}
    \centering
    \includegraphics[width = 0.85\textwidth]{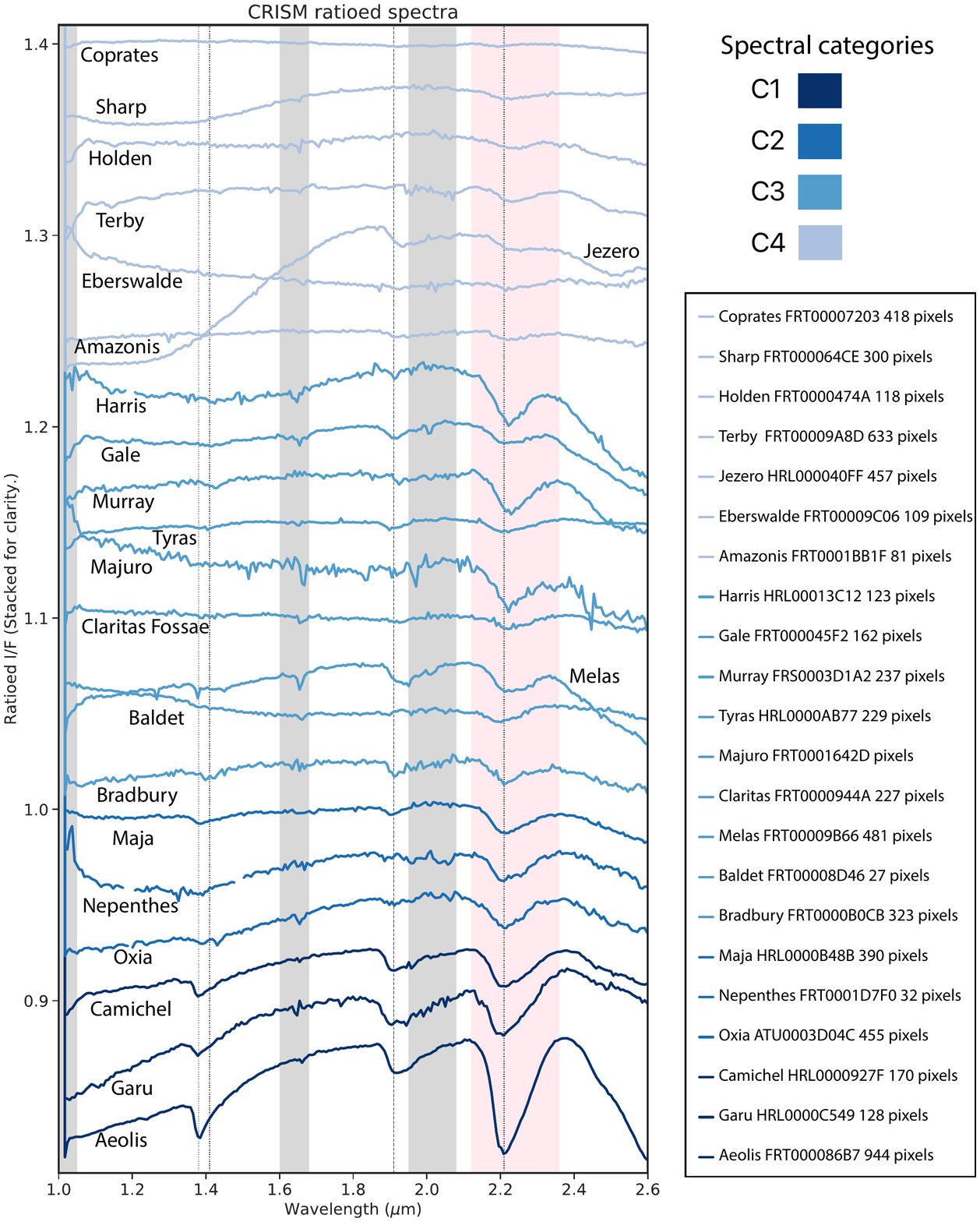}
    \caption{Type silica spectra from all sites with alluvial fan or delta deposits, excluding the silica deposits with ambiguous context. }
    \label{fig:A1}
\end{figure}

\begin{figure}
    \centering
    \includegraphics[width = 0.85\textwidth]{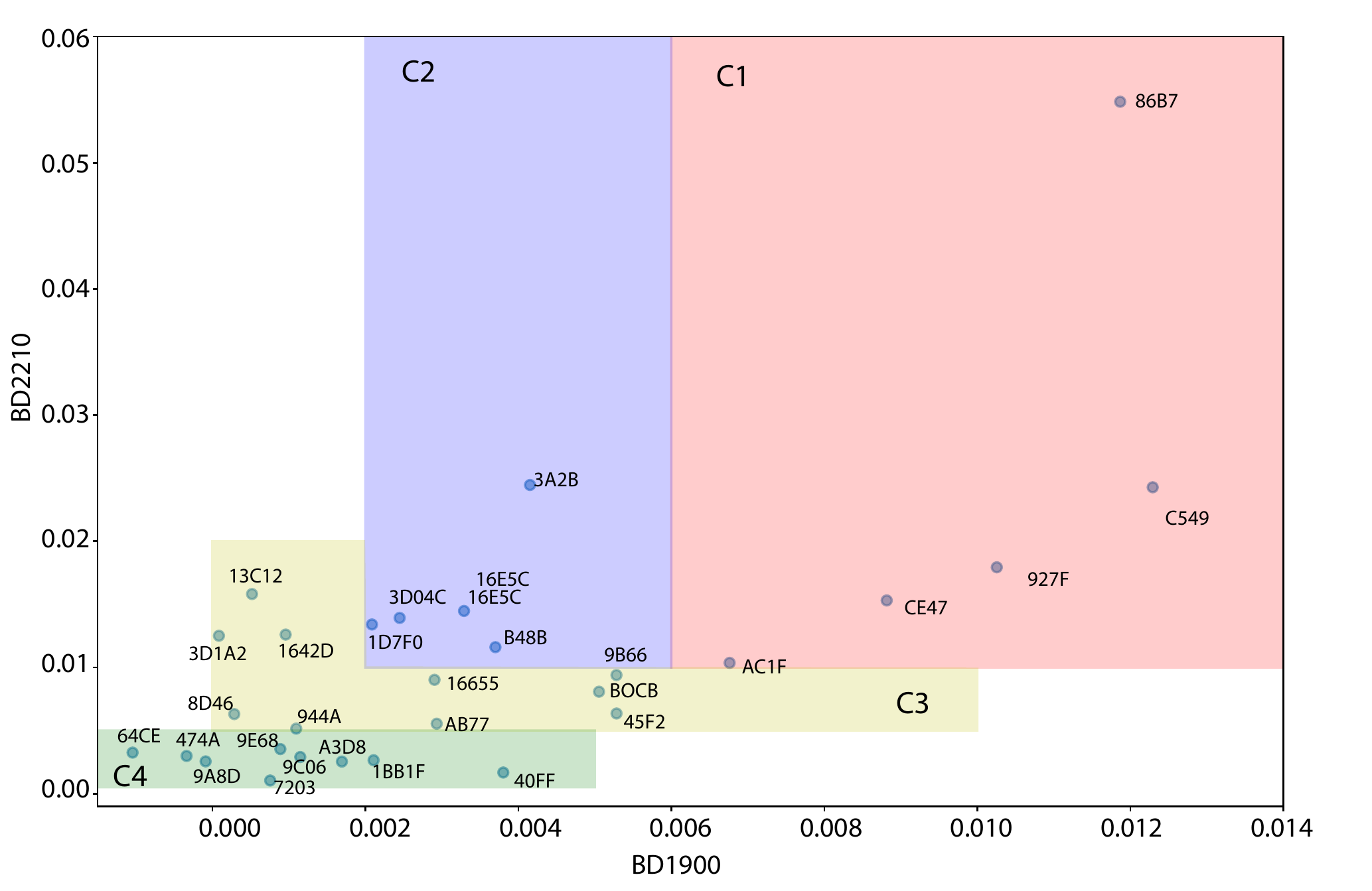}
    \caption{Classification of spectral categories based on classic band depth formula. Limits for each category are given in Table S3. }
    \label{fig:A2}
\end{figure}

\begin{figure}
    \centering
    \includegraphics[width = 0.85\textwidth]{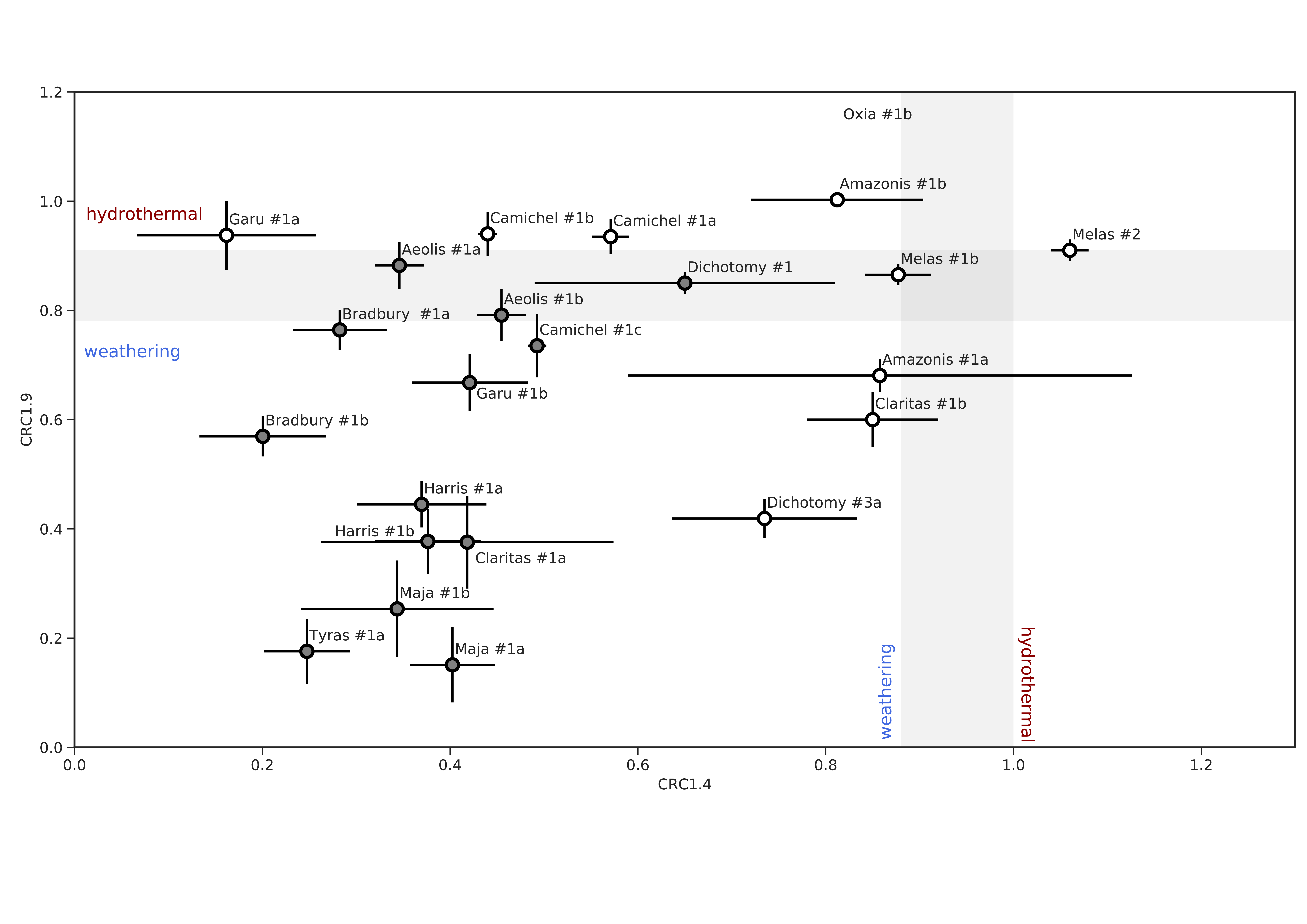}
    \caption{The CRC parameters of silica spectra compared to previous lab analysis and other Mars locations \citep{pineau_toward_2020}. Plots of spectral parameters showing differentiating features between silica formation in hydrothermal vs. low-temperature weathering systems. Here the CRC 1.9 parameter is highly sensitive to the CO2 atmospheric residual. CRC 1.4 parameter could be affected by the low signal-to-noise ratio at this wavelength. Data points of lower confidence levels are shown with empty circles. The data and information on the spectra used to make these plots are given in Table S6. }
    \label{fig:A3}
\end{figure}

\begin{figure}
    \centering
    \includegraphics[width = 0.65\textwidth]{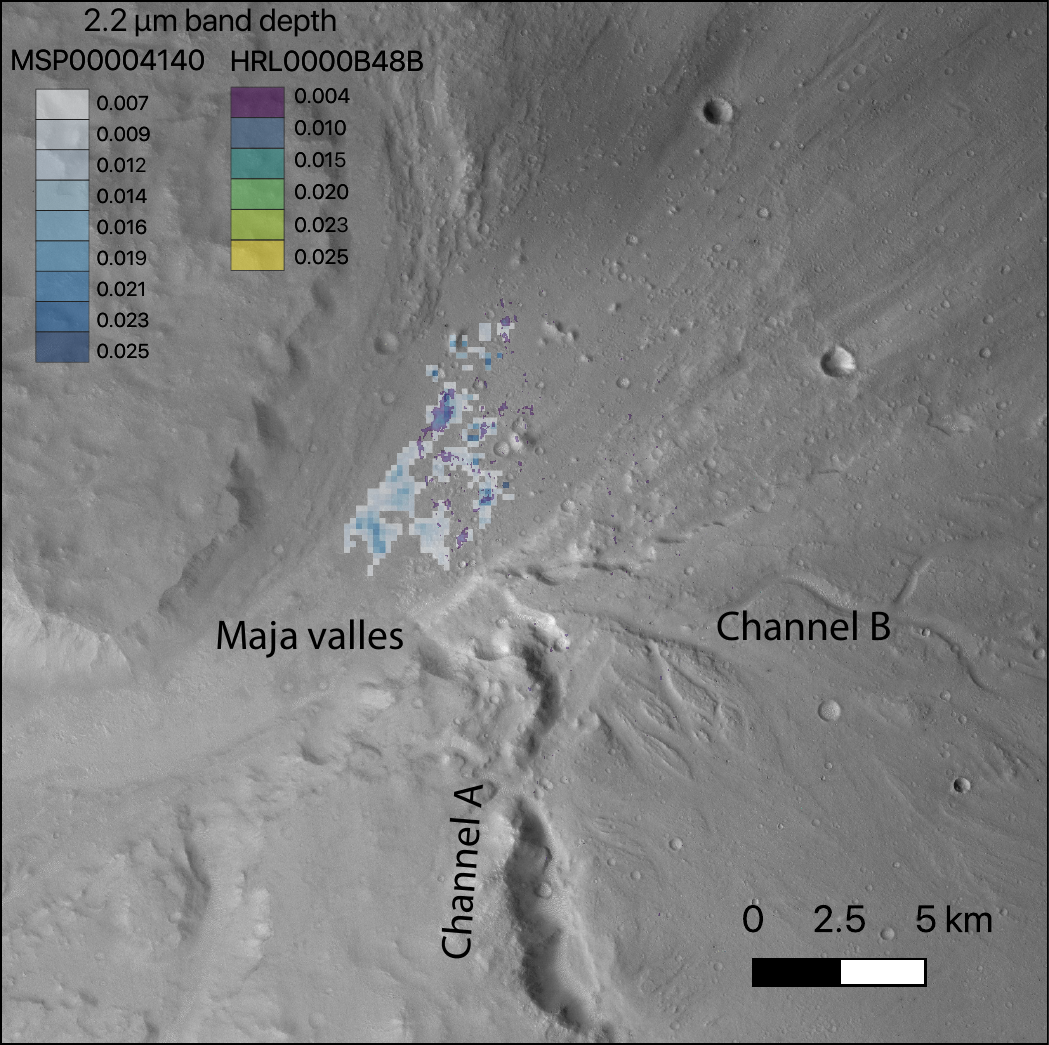}
    \caption{Hydrated silica detection at the terminal deposit of Maja Valles. Silica detection is not directly correlated with a specific fan/delta morphology but is likely the result of past fluvial activities. Other than the main outflow channel (Maja Valles) in which the silica is found, there are two smaller, likely more recent channels (A, B), which may also have contributed. Both the multispectral tile (MSP00004140) and hyperspectral half-resolution tile (HRL0000B48B) are found with this signature. Here band depths of these two images are highlighted using different color scales. }
    \label{fig:A4}
\end{figure}

\begin{figure}
    \centering
    \includegraphics[width = 0.75\textwidth]{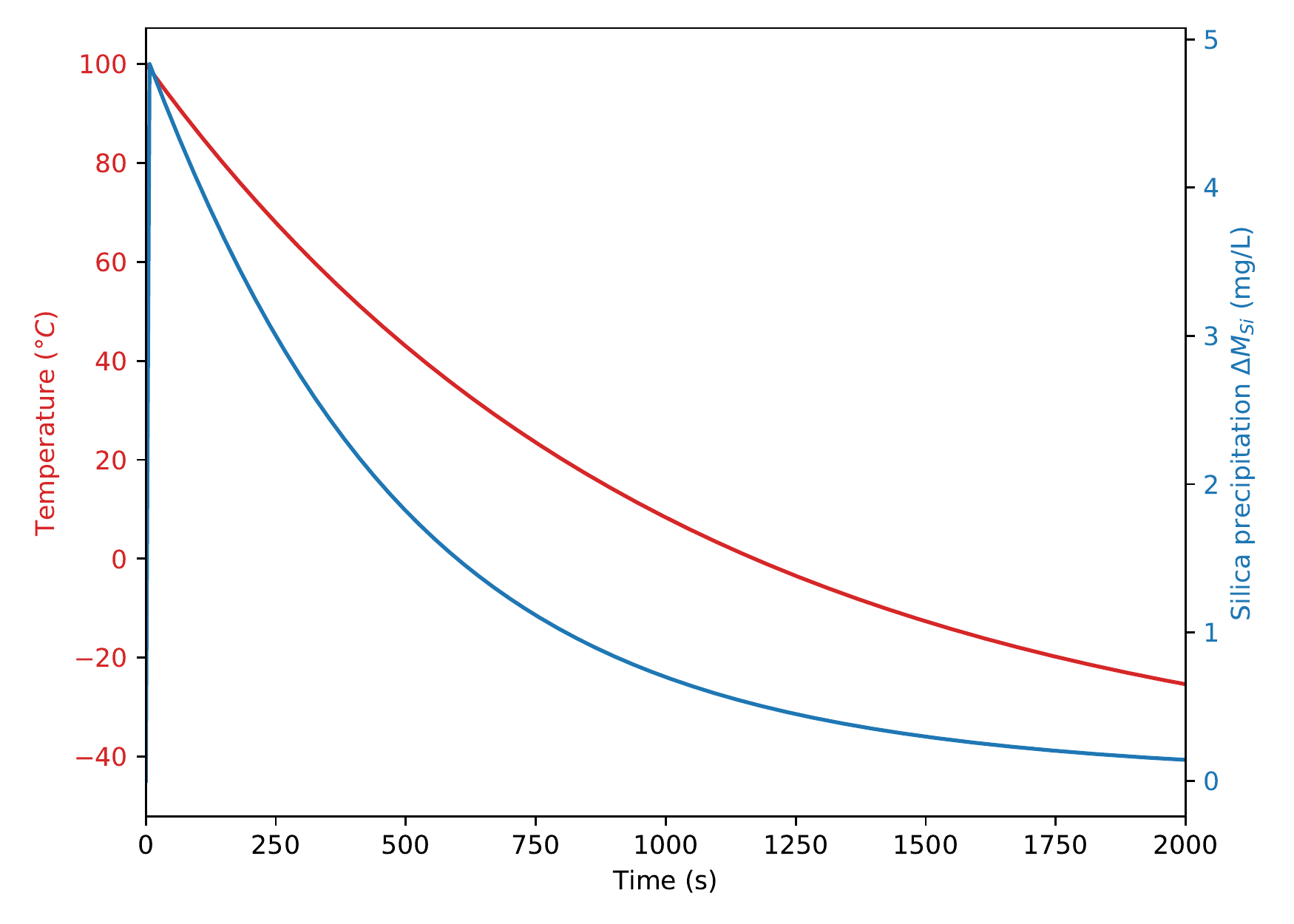}
    \caption{A simplistic model of the predicted amount of silica precipitation prediction in an equilibrium water system in contact with Mars surface condition (T=228K). The amount of silica precipitated per second is expected to decay exponentially as the temperature decays with time.  }
    \label{fig:A5}
\end{figure}

\begin{figure}
    \centering
    \includegraphics[width = 0.75\textwidth]{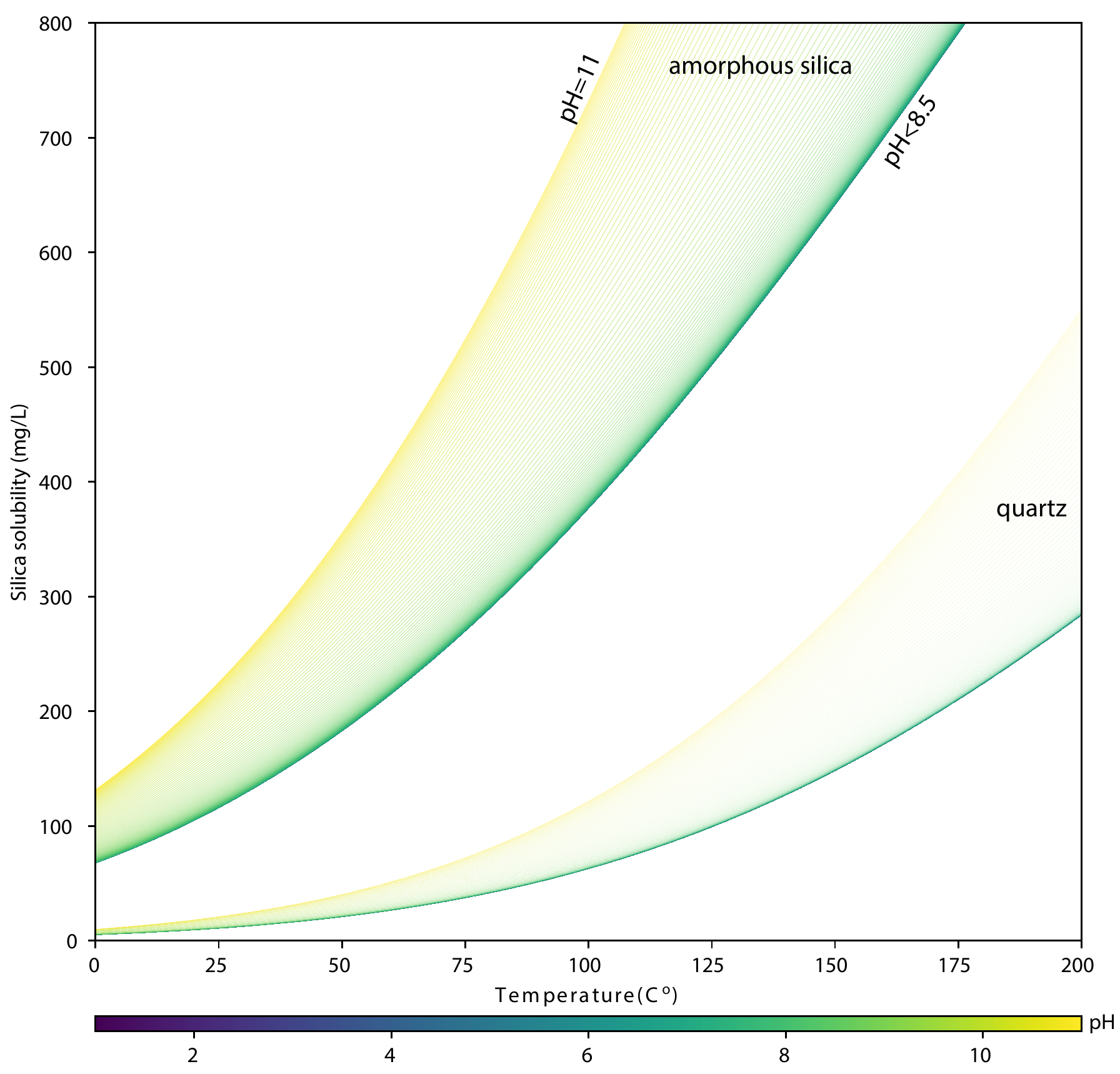}
    \caption{Calculated solubility ranges of silica-based on observation compared with silica solubility curve correlation with temperature and pH. The silica dissolved in water remains mostly monomeric silicic acid until pH = 9-10. The silica solubility is calculated based on the equilibrium constant of the first ionic dissociation of silicic acid.}
    \label{fig:A6}
\end{figure}

\begin{figure}
    \centering
    \includegraphics[width = 0.75\textwidth]{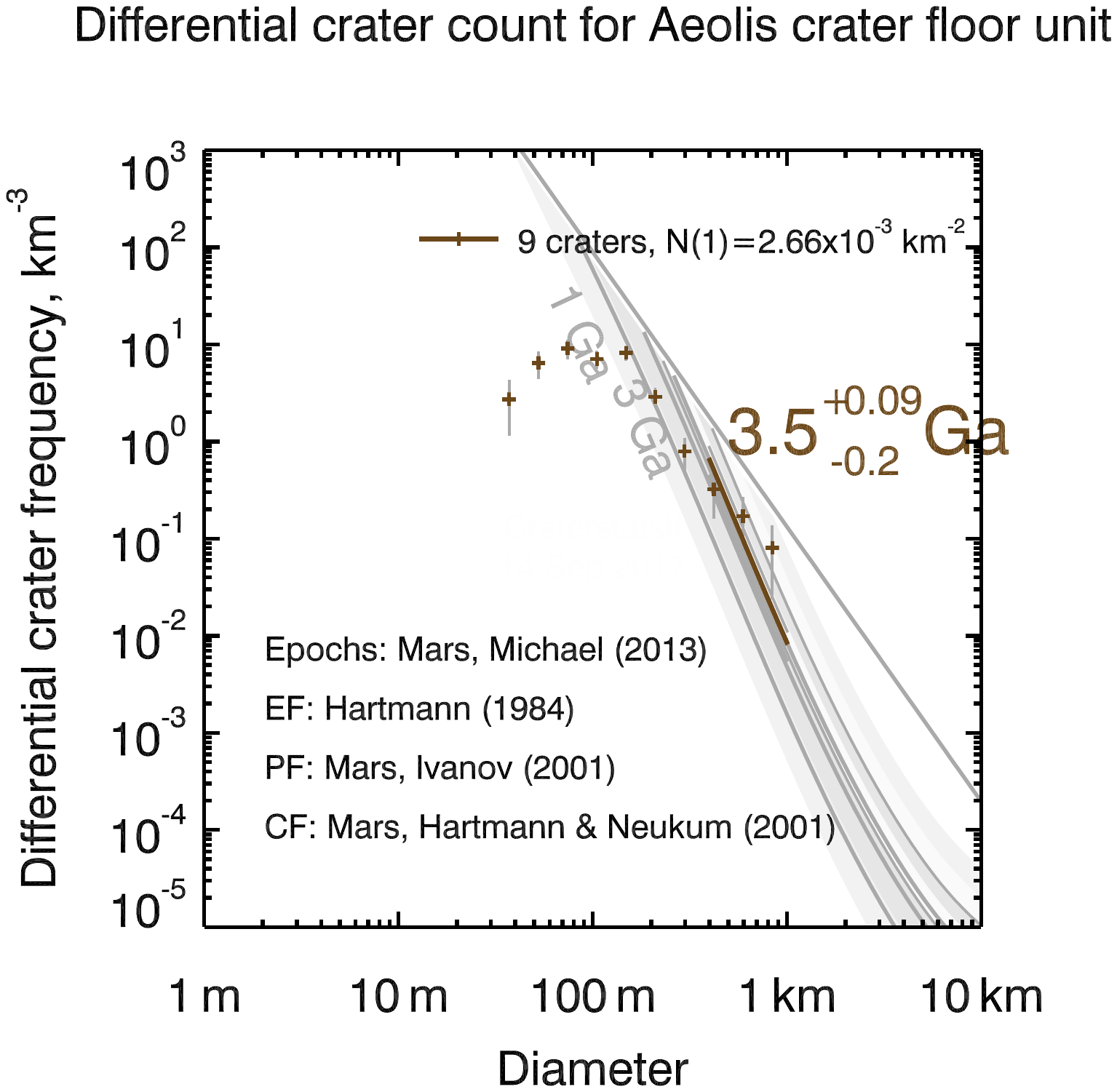}
    \caption{Crater count ages of the Aeolis fan crater floor unit.}
    \label{fig:A7}
\end{figure}

\section{Figure Set 3.}
\begin{figure}
    \centering
    \includegraphics[width=0.75\textwidth]{Figure3_r1-01.jpg}
    \includegraphics[width=0.75\textwidth]{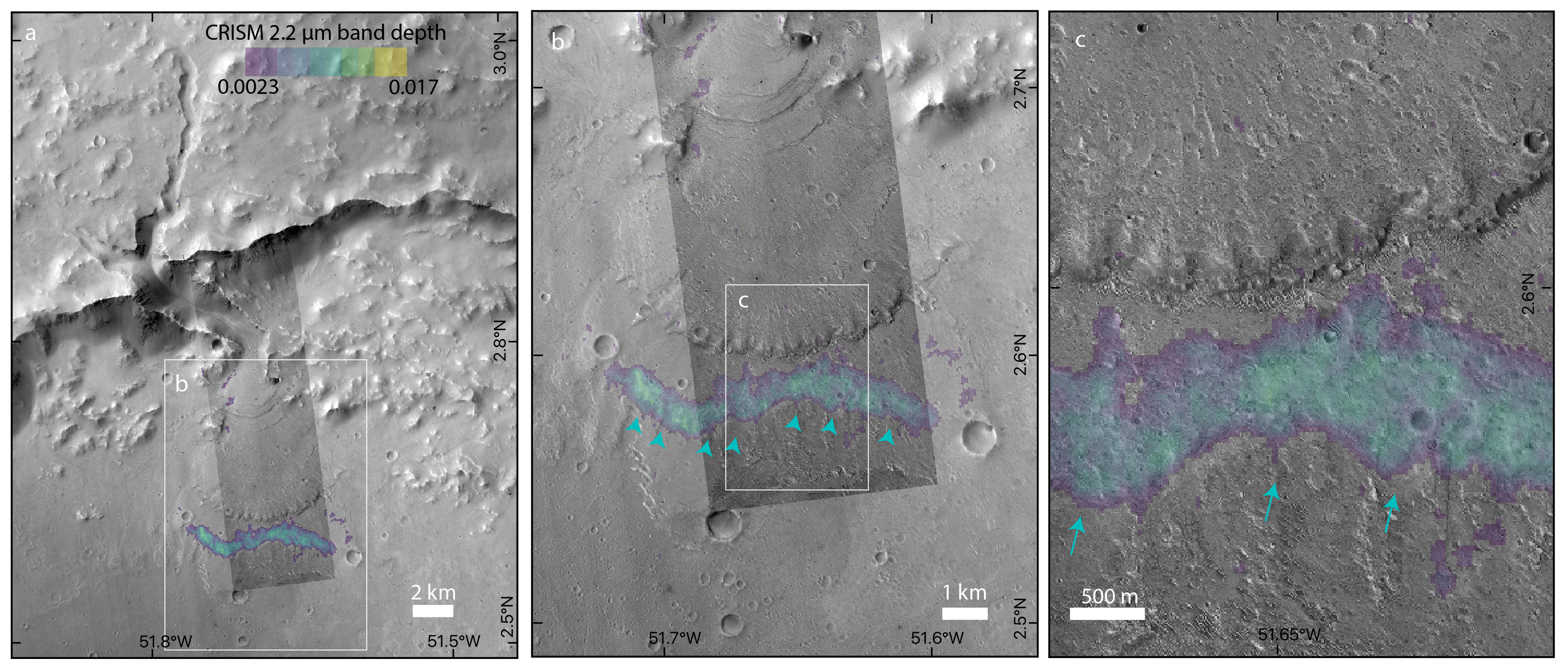}
    \includegraphics[width=0.75\textwidth]{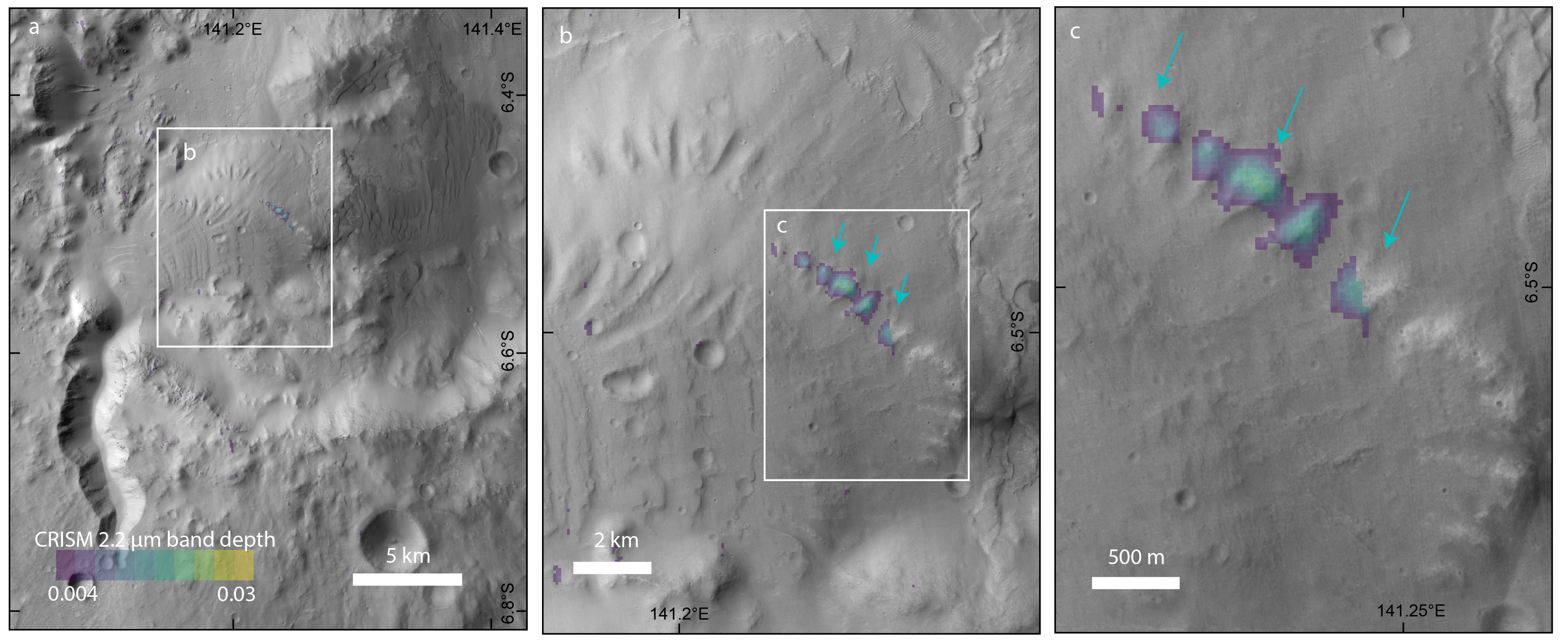}
    \caption{CRISM overlay of silica detection on Aeolis, Camichel, Garu}
    \label{fig:3.1}
\end{figure}

\begin{figure}
    \centering
    \includegraphics[width=0.75\textwidth]{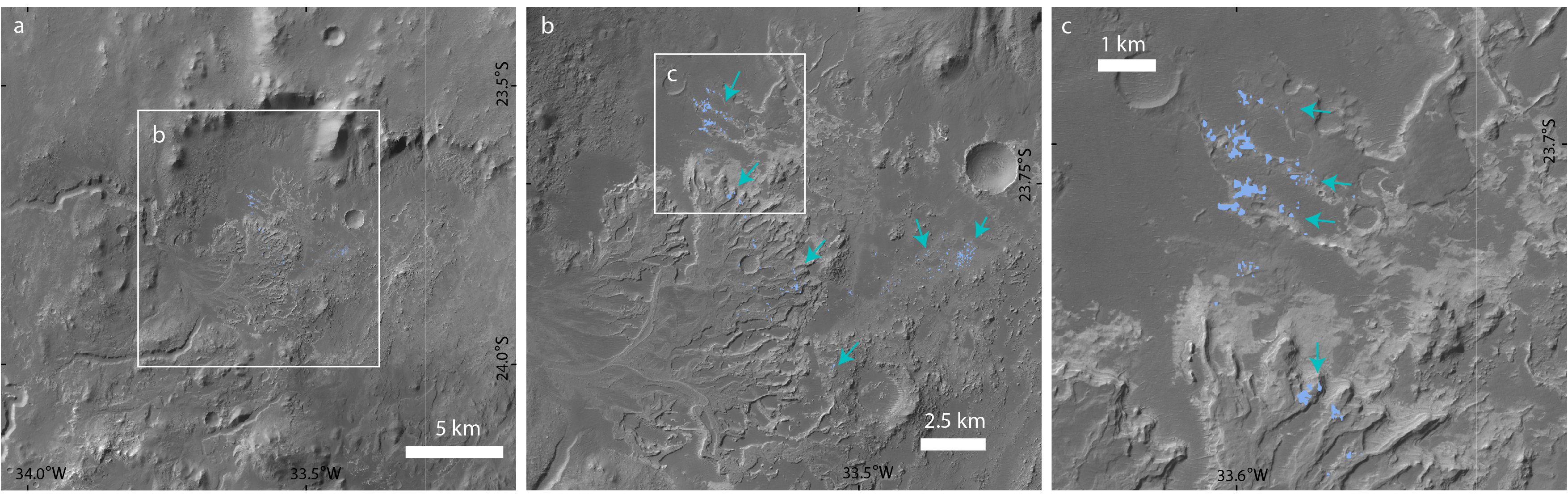}
    \includegraphics[width=0.75\textwidth]{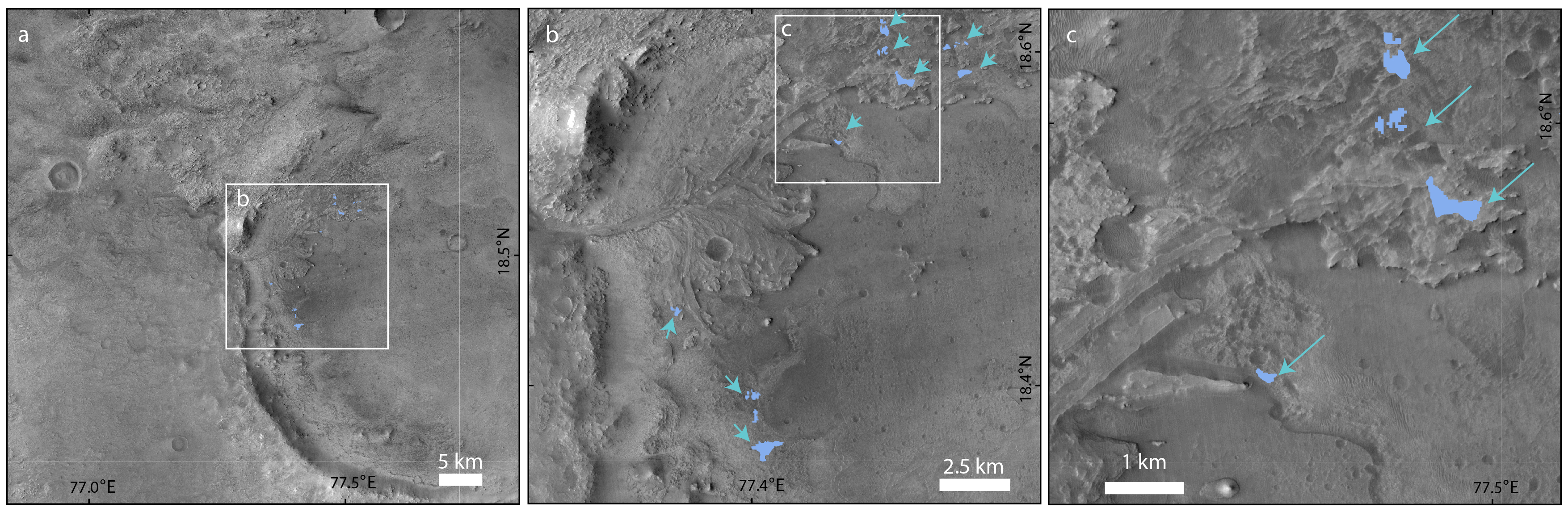}
    \includegraphics[width=0.75\textwidth]{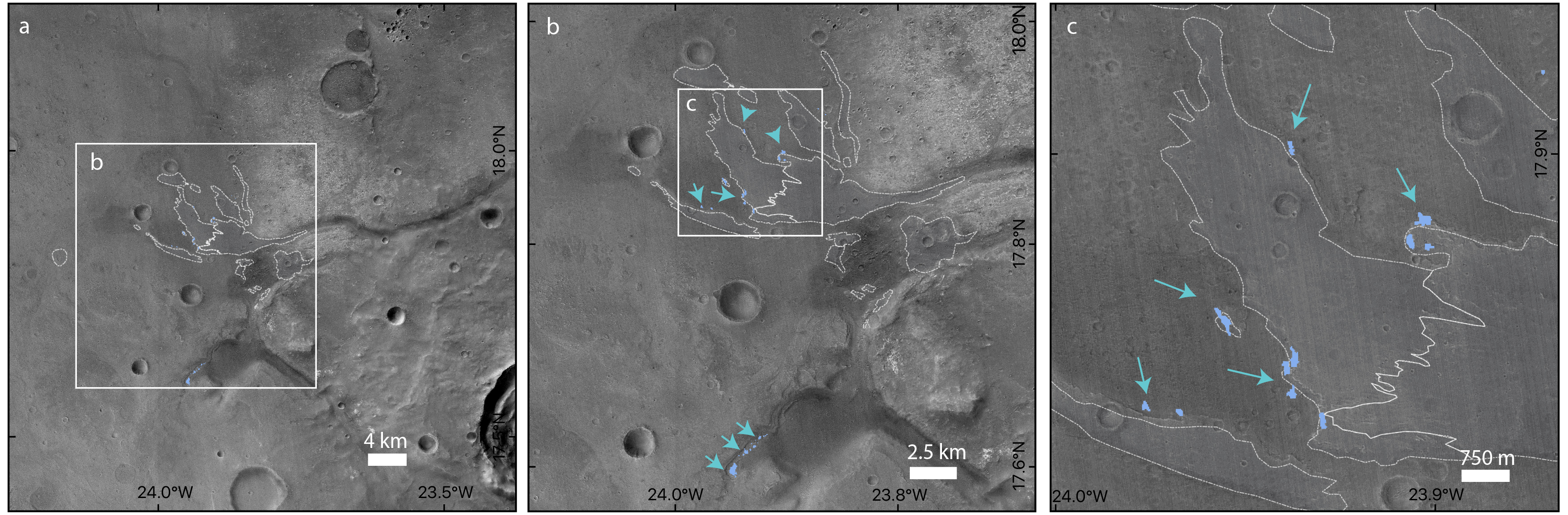}
    \caption{CRISM overlay of silica detection on Eberswalde, Jezero, Oxia}
    \label{fig:3.2}
\end{figure}

\section{Supplementary tables: Table S1-S7 (See supporting data)}

\end{document}